\documentclass{aa}

\usepackage{graphicx}
\usepackage{txfonts}
\usepackage[percent]{overpic} %pics in tables
\usepackage{pdflscape}
\usepackage{amsmath}
\usepackage{comment}
\usepackage{enumitem}
\usepackage{booktabs} %Handling tables
\usepackage{multirow} %Handling tables
\usepackage{makecell} %Handling tables
\usepackage[export]{adjustbox} %To align figures also on the left/right

\newcommand{\Mj}{\rm M_{Jup}}
\newcommand{\phip}{\phi_{p}}

\newcommand{\discminer}{\textsc{discminer}}
\newcommand{\twCOfull}{$^{12}$CO\,$J=2-1$}
\newcommand{\twCO}{$^{12}$CO}
\newcommand{\polaris}{\textsc{polaris}}
\newcommand{\sfdmodels}{\textsc{sf{\scriptsize 3}dmodels}}
\newcommand{\fargo}{\textsc{fargo{\scriptsize 3}d}}

\begin{document}

   \title{The Disc Miner I: A statistical framework to detect and quantify kinematical perturbations driven by young planets in discs} 
    \titlerunning{The Disc Miner I}

   \subtitle{}

   \author{A. F. Izquierdo\inst{1}\fnmsep\inst{2}
          \and
          L. Testi\inst{1}\fnmsep\inst{3}\fnmsep\inst{4}
          \and
          S. Facchini\inst{1}
          \and
          G. P. Rosotti\inst{2}\fnmsep\inst{5}
          \and
          E. F. van Dishoeck\inst{2}\fnmsep\inst{6}
          }

   \institute{European Southern Observatory, Karl-Schwarzschild-Str. 2, 85748 Garching bei München, Germany\\
              \email{andres.izquierdo.c@gmail.com}
         \and
             Leiden Observatory, Leiden University, P.O. Box 9513, NL-2300 RA Leiden, The Netherlands
         \and 
             Excellence Cluster Origin and Structure of the Universe, Boltzmannstr. 2, 85748 Garching bei München, Germany
         \and
             INAF – Osservatorio Astrofisico di Arcetri, Largo E. Fermi 5, 50125 Firenze, Italy
         \and
             School of Physics and Astronomy, University of Leicester, Leicester LE1 7RH, UK
         \and 
             Max-Planck-Institut für extraterrestrische Physik, Gießenbachstr. 1 , 85748 Garching bei München, Germany
             }

   \date{Received 10 March 2021 / Accepted 21 April 2021}

% \abstract{}{}{}{}{} 
% 5 {} token are mandatory
 
  \abstract
  % context heading (optional)
  % {} leave it empty if necessary  
   {The study of disc kinematics has recently opened up as a promising method to detect unseen planets. However, a systematic, statistically meaningful analysis of such an approach remains missing in the field.}
  % aims heading (mandatory)
   {The aim of this work is to devise an automated, statistically robust technique to identify and quantify kinematical perturbations induced by the presence of planets in a gas disc, and to accurately infer the location of the planets. 
   }
  % methods heading (mandatory)
   {We produced hydrodynamical simulations of planet--disc interactions with different planet masses, namely 0.3, 1.0, and 3.0\,$\Mj$, at a radius of $R_p=100$\,au in the disc, and performed radiative transfer calculations of CO to simulate observables for a disc inclination of $-45^\circ$, and for 13 planet azimuths. We then fitted the synthetic data cubes with a Keplerian model of the channel-by-channel emission using the \discminer{} package. 
   Lastly, we compared the synthetic cubes with the best-fit model to: extract deviations from Keplerian rotation; and quantify both large-scale and localised intensity, line width, and velocity fluctuations triggered by the embedded planets and provide strong constraints on their location in the disc. We assess the statistical significance of the detections using the peak and variance of the planet-driven velocity fluctuations. 
   }
  % results heading (mandatory)
   {Our findings suggest that a careful inspection of line intensity profiles to analyse gas kinematics in discs is a robust method to reveal embedded, otherwise unseen planets, as well as the location of gas gaps. We claim that a simultaneous study of line-of-sight velocities and intensities is crucial to understanding the origin of the observed velocity perturbations. In particular, the combined contribution of the upper and lower emitting surfaces of the disc plays a central role in setting the observed gas velocities. This joint effect is especially prominent and hard to predict at the location of a gap or cavity, which can lead to artificial deviations from Keplerian rotation depending on how the disc velocities are retrieved.
   Furthermore, regardless of their origin, gas gaps alone are capable of producing kink-like features on intensity channel maps, which are often attributed to the presence of planets. 
   Our technique, based on line centroid differences, takes all this into account to capture only the strongest, localised, planet-driven perturbations. It does not get confused by axisymmetric velocity perturbations that may result from non-planetary mechanisms. The method can detect all three simulated planets, at all azimuths, with an average accuracy of $\pm3^\circ$ in azimuth and $\pm8$\,au in radius. As expected, velocity fluctuations driven by planets increase in magnitude as a function of the planet mass. Furthermore, owing to disc structure and line-of-sight projection effects, planets at azimuths close to $\pm45^\circ$ yield the highest velocity fluctuations, whereas those at limiting cases, $0^\circ$ and $\pm90^\circ$, drive the lowest. The observed peak velocities typically range within 40$-$70\,m\,s$^{-1}$, 70$-$170\,m\,s$^{-1}$, and 130$-$450\,m\,s$^{-1}$ for 0.3, 1.0, and 3.0\,$\Mj$ planets, respectively. Our analysis indicates that the variance of peak velocities is boosted near planets because of organised gas motions prompted by the localised gravitational well of planets. We propose an approach that exploits this velocity coherence to provide, for the first time, statistically significant detections of localised planet-driven perturbations in the gas disc kinematics.
   }
  % conclusions heading (optional), leave it empty if necessary 
   {}


   \maketitle

%***********************
\section{Introduction}

In order to detect young planets, it is imperative to understand the footprints they leave on protoplanetary discs, their place of  formation. Recent observations in concert with theoretical efforts suggest that dust substructures, mostly rings and gaps but also cavities, spirals, and asymmetric features, are possibly ubiquitous in protoplanetary discs \citep{alma+2015, isella+2016, perezL+2016, long+2018, andrews+2018}. In multiple cases, embedded planets may play a key role in shaping some of these substructures observed in infrared and (sub)millimetre wavelengths \citep[see e.g.][]{benisty+2015, dipierro+2015, pinilla+2018, zhang+2018, ubeira+2019, facchini+2020}. However, planet--disc interactions are far from being the only driving mechanism behind dust signatures in young discs. Magnetic, hydrodynamic, and gravitational instabilities can also lead to dust substructure \citep{armitage+2011, andrews+2020}, meaning that looking at the dust emission alone is generally insufficient to unambiguously claim the presence of planets. On top of that, thermal and accretion emission from young planets is hard to detect through direct imaging, 
whose range of action is currently narrowed to massive planets and low-dust-extinction scenarios \citep{testi+2015, sanchis+2020}. To date, PDS 70 is the only system in which forming planets have been convincingly detected by direct imaging \citep[PDS 70b and PDS 70c,][]{keppler+2018, haffert+2019}.

Luckily, not only dust but also gas stores valuable information that can help disentangle the physical and chemical processes at work (and often coupled) in planet-forming discs \citep{bruderer+2012, henning+2013, dutrey+2014}. The presence of deep gas cavities, smaller than those in dust at (sub)mm wavelengths, is a clear diagnostic of the disc interaction with planet and stellar companion(s) \citep{bruderer+2014, perez+2015a, vandermarel+2015, vandermarel+2016a}. On the other hand, the rich molecular gas disc inventory has been meticulously examined over recent years to reveal density and temperature structure, as well as elemental abundances in a number of objects \citep{pietu+2007, rosenfeld+2013, williams+2014, miotello+2016, dutrey+2017, pinte+2018a, dullemond+2020, rosotti+2020, teague+2020a, facchini+2021}. These are all crucial pieces in the vast puzzle of planet formation \citep{benz+2014, johansen+2017, oberg+2021}. 
Moreover, spectral lines from molecules provide a useful window onto gas velocities, and can be {`mined'} with appropriate modelling to understand the mechanisms driving the disc dynamics \citep[see e.g.][]{rosenfeld+2013, price+2018, teague+2019nat, woelfer+2021, paneque-carreno+2021}, and in consequence, to better constrain the presence of planets. 

It is well known from previous theoretical works that hydrodynamical and gravitational interactions between forming planets and gas discs produce particular signatures in molecular line observations \citep{perez+2015, perez+2018}. Line Doppler shifts due to localised deviations from Keplerian rotation are expected in circumplanetary material and along spiral wakes launched from the planet. Furthermore, embedded planets carve density gaps which induce pressure gradients in the gas disc, producing azimuthally extended non-Keplerian velocities at the edges of the gaps \citep[see e.g.][]{diskdynamics+2020}. In this context, and supported by the high angular and spectral resolution offered by the Atacama Large (sub)Millimeter Array (ALMA), many are increasingly turning their attention to spotting the kinematical clues left by planets in protoplanetary discs. 

More specifically, \citet{pinte+2018b, pinte+2019} claimed kinematical detections of giant planets embedded in the discs around HD\,163296 and HD\,97048 by empirical comparison of planet-disc hydrodynamic simulations capable of producing kink-like features similar to those observed in CO intensity channel maps. A handful of new kinks in \twCO{} were reported later by \citet{pinte+2020} but most of them await confirmation because of limitations in signal-to-noise ratio and spectral resolution. The kink velocity can be linked to the driving planet mass through simple relationships as recently theorised by \citet{rabago+2021} and \citet{bollati+2021}, yet both of these latter studies omit radiative transfer effects. On the other hand, \citet{teague+2018a} proposed the presence of two other giant planets in shorter orbits around HD\,163296 by looking at azimuthally symmetric gas velocity deviations from Keplerian rotation due to pressure gradients driven by planet-induced gaps. However, this method is limited by the fact that gaps opened by other mechanisms would drive similar kinematical signatures in the disc \citep[see e.g.][]{rabago+2021}.
Using rotation curves on the disc of HD\,100546, \citet{casassus+2019} detected a localised `Doppler flip' in \twCO{} reminiscent of the velocity perturbations expected along spiral wakes induced by a planet. Nevertheless, although pivotal, these studies are still source-specific and some lack statistical significance. In particular, the presence of kinks is currently assessed by visual inspection of channel maps, which works fine in some clear cases, but does not allow for an estimate of the significance of the detection, implying that the method can be misleading in less apparent cases.

In this paper, we introduce a statistical framework to overcome these limitations and to robustly detect localised perturbations due to unseen planets using molecular line observations. The technique is also applied to systematically extract {observable} velocity perturbations driven by different planet masses at a number of azimuths in a synthetic disc. 
The outline of the work is as follows. Section \ref{sec:simulations_rt} describes the planet--disc interaction simulations and synthetic observations used throughout the work. Section \ref{sec:discminer} introduces the \discminer{} package and how we use it to fit Keplerian intensity channel maps on the synthetic observations. Section \ref{sec:results} presents the extraction and analysis of non-Keplerian velocities, as well as the statistical framework designed to infer the location of planets as a function of planet mass and azimuth in the disc. We expand on the advantages of our method and discuss the observable signatures of planets in Section \ref{sec:discussion}, and summarise the main results in Section \ref{sec:conclusions}.   

%***********************
\section{Hydrodynamic simulations and radiative transfer} \label{sec:simulations_rt}

%-----------------------------------------------------------------
   \begin{figure*}
   \centering
   \includegraphics[width=1\textwidth]{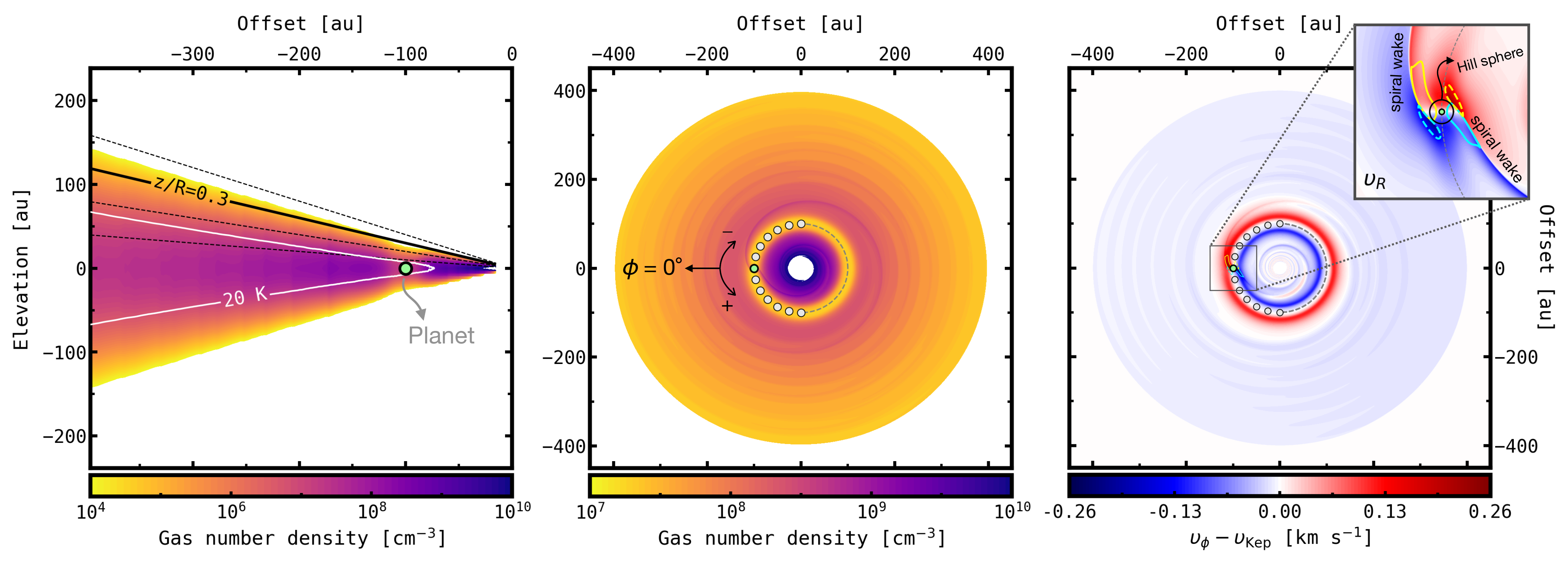}
      \caption{\textit{Left panel}: Edge-on gas number density of one of our planet--disc simulation snapshots (1.0\,$\Mj$, $\phi_{p}=0^\circ$). The white contour encloses the CO freeze-out region ($T<20$\,K). The black lines correspond to $z/R$ = 0.1, 0.2, 0.3, 0.4 scale heights, with the solid line being the threshold adopted for photodissociation. 
      \textit{Middle panel}: Face-on view of the gas number density in the midplane of the disc. \textit{Right panel}: Azimuthal (main panel) and radial (zoom-in) deviations from Keplerian velocity. The solid and dashed contours in the zoomed-in panel are $\pm$60 per cent peak azimuthal and peak radial perturbations, respectively, and illustrate that both components do not necessarily overlap and in turn contribute independently to the observed peak velocity perturbations. Also shown is the planet Hill sphere, with a radius of 6.8\,au. The green circle indicates the current position of the planet, and the grey circles are additional planet azimuths explored in this work.
              }
         \label{fig:dens+temp}
   \end{figure*}
%-----------------------------------------------------------------   

To simulate velocity perturbations triggered by planet--disc interactions, we use the multifluid open source code \fargo{} \citep{benitez+2016} to solve the hydrodynamic evolution of a gas disc, which responds to thermodynamic variables via the Navier-Stokes and continuity equations, and to the gravitational potential of point-like sources.

Assuming viscous accretion \citep{lyndenbell+1974, hartmann+1998}, we initialise the simulations with a gas disc surface density profile of the form $\Sigma_{\rm gas}(R)=\Sigma_c (R/R_c)^{-\gamma} \exp[-(R/R_c)^{2-\gamma}]$, with $R_c=100$\,au as the characteristic radius, $\Sigma_c=3.0$\,g\,cm$^{-2}$ 
 the density normalisation at $R_c$, and $\gamma=1.0$ the viscous power-law exponent. We adopt a uniform kinematic viscosity of $\alpha=10^{-3}$, and a 1\,M$_\odot$ point-like source at the centre of the mesh. The mesh spans from $-\pi$ to $\pi$ in $\phi$, 15 to 700\,au in $R$, and $-$240 to 240\,au in $z$. The number of cells is 2048, 1250, and 149, respectively, evenly spaced in $\phi$ and $z$, and logarithmically spaced in $R$. The inner and outer radii are assumed to be $R_{\rm in}=15$\,au and $R_{\rm out}=400$\,au, yielding a gas mass of 0.01\,M$_\odot$. 

We ran three independent 2D simulations varying the mass of the embedded planet (0.3, 1.0, and 3.0\,$\Mj$) at a fixed radial location $R_p=100$\,au, and let them evolve for 1000 orbits to reach a steady state. Even though the mesh of our simulations is not adaptive, its spatial resolution guarantees that the Hill radius of the planet is resolved (e.g. with $\sim$\,22 cells for the 1.0\,$\Mj$ planet). Also, the planets are not treated as sink particles, so they are not accreting material. Their gravitational potential is smoothed with 0.6 of the disc scale height, which is the closest 2D approximation of 3D gravity \citep{mueller+2012}. On the other hand, we do not find fully developed vortices in any of the simulations at steady state. This is a consequence of the kinematic viscosity being high enough to efficiently quench vortices \citep[see e.g.][]{fu+2014, hammer+2017}.

To extend the simulations to three dimensions, we assume a cylindrical velocity field. Because of the planet gravity, it is not straightforward to calculate the rotation velocity at different heights in the disc based on the midplane velocities and the central force of the star only, let alone the radial and vertical velocity components. Full 3D simulations are required to self-consistently tackle this. On the other hand, we consider hydrostatic equilibrium to compute the gas volume density along the vertical direction, $\rho_{\rm gas}(R, z) = \left(\Sigma_{\rm gas}(R)/\sqrt{2\pi}H\right)\exp[-0.5(z/H)^2]$, where $H(R)=H_0(R/100\,{\rm au})^{\psi}$ is the scale height of the disc, with a normalisation $H_0=6.5$\,au, and a flaring index $\psi=1.25$. Figure \ref{fig:dens+temp} shows edge-on and face-on slices of the resulting gas number density and intrinsic deviations from Keplerian rotation for the 1.0\,$\Mj$ simulation.

Based on the simulated gas densities and velocities, we perform radiative transfer calculations to obtain synthetic emission maps of \twCOfull{} assuming a parametrised CO abundance (see below). Velocity binning, line-of-sight projection and optical depth effects are therefore considered in the analysis. These are unavoidable factors that must be taken into account if one is to study any observable, especially when it is expected to be faint and confined to small scales as is generally the case for planet-driven perturbations \citep{perez+2018}. In real data, additional sources of uncertainty such as noise in the signal, beam smearing, and non-linear artefacts from image reconstruction algorithms are also important \citep[see e.g.][]{diskdynamics+2020} and will be explored in future releases of the Disc Miner series, which are more focused on observations.

We use the \textsc{rt} modules of the \sfdmodels{} package \citep{izquierdo+2018} to bridge our \fargo{} simulation snapshots with the \polaris{} radiative transfer code \citep{reissl+2016}. We first run \polaris{} to compute the three-dimensional dust temperature, which is done by propagating photon packages semi-randomly from a T Tauri star at the centre of the disc, with a luminosity of 1.9\,L$_\odot$ and a photosphere at an effective temperature of 4000\,K. We assume a standard Mathis radiation field at a galactocentric distance of 10\,kpc as an external source of radiation \citep{mathis+1983}. The disc is located 100\,pc away from the observer and inclined at $-45^\circ$ with respect to the plane of the sky, with the north half being the side closest to the observer. For simplicity, the position angle of the disc was fixed at $0^\circ$. The dust model consists only of silicate grains with an intrinsic density of 3.5\,g\,cm$^{-3}$ \citep[with optical properties from][]{weingartner+2001}. We use a standard ISM gas-to-dust mass ratio of 100, and a grain size distribution $n(a)=a^{-3.5}$, between $a_{\rm min}=5$\,nm and $a_{\rm max}=1\,\rm mm$. We consider a direct conversion of temperatures $T_{\rm gas}=T_{\rm dust}$, which is a good assumption at scale-heights $z/R<0.3$ where our analysis takes place \citep[see e.g.][]{kamp+2004, jonkheid+2004, miotello+2014}. Second, we use \sfdmodels{} again to read the output temperatures and add simplified chemical processes such as CO freeze-out on dust grains, where $T_{\rm dust}<20$\,K, and photodissociation, at scale heights $z/R>0.3$.
We adopt a \twCO{} abundance of $5\times10^{-5}$ in the gas phase, and $5\times10^{-11}$ in freeze-out and photodissociated regions. Lastly, \sfdmodels{} provides \polaris{} with the gas velocity, temperature, and the processed CO abundance distribution, which this time are used to compute the ray-traced intensity channel maps of the simulations. 

\begin{figure*}
   \begin{center}  
   \includegraphics[width=0.94\textwidth]{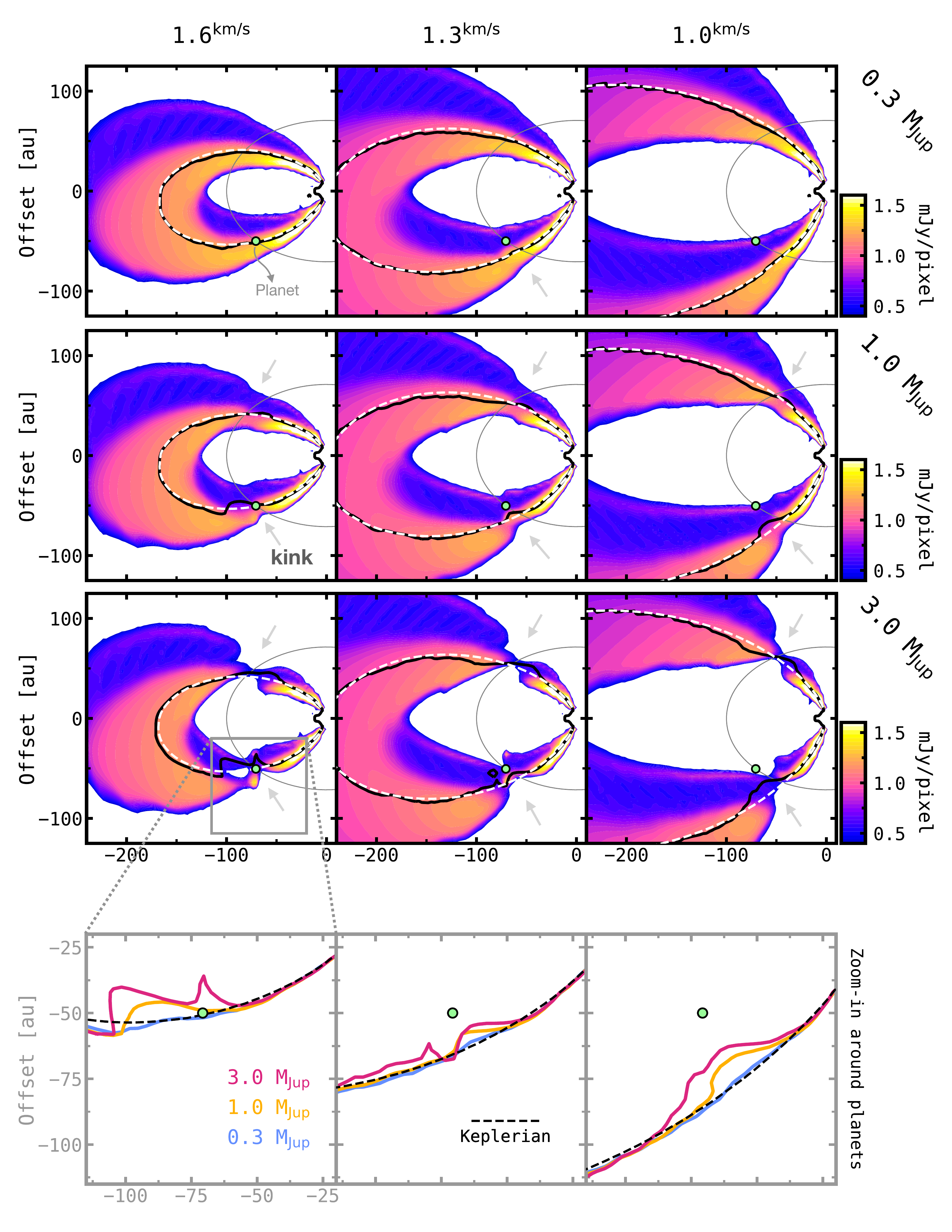}
      \caption{Selected \twCOfull{} synthetic channel maps for the three simulation snapshots (0.3, 1.0 and 3.0\,$\Mj$, from top to bottom), with the planets at $R_p=100$\,au radius, and $\phip{}=45^\circ$ azimuth, marked as green circles. The disc is inclined at $-45^\circ$ with respect to the plane of the sky. The thin grey line is the projected circular orbit of the planet. The solid lines are centroid velocity contours extracted from the simulations at the velocity channel indicated on the top header; the dashed lines show the same but for the Keplerian best-fit model. Small arrows indicate kink-like features identified by visual inspection. In the bottom row a zoom-in around planets is shown for better comparison of centroid velocities and their deviation from Keplerian rotation as a function of planet mass.
              }
         \label{fig:channel_maps}
         
\end{center}
   \end{figure*}

In order to consider projection effects on the planet-driven perturbations, we ran the ray-tracing 13 times per planet, each time rotating the input gas properties along the vertical axis of the disc from --90$^\circ$ to 90$^\circ$, in steps of 15$^\circ$, so that the planets can be observed at 13 different azimuths in the posterior analysis (see Fig. \ref{fig:dens+temp}). As such, we end up with 39 position-position-velocity (ppv) cubes, each with 101 channel maps ranging from $-$5 to 5\,km\,s$^{-1}$, taking a channel width of 0.1\,km\,s$^{-1}$. We adopt a pixel size of 3.1\,au which translates to 31\,mas at 100\,pc. A selection of these channel maps is presented in Figure \ref{fig:channel_maps}, where we also compare some of the kinematical features triggered by the three planets in our sample. There are many kinks evident to the eye in the channel maps, especially around massive planets. However, we note that the gap alone can also produce kink-like features, suggesting that empirical methods that rely on visual inspection are prone to false positive detection of planets. Conversely, as explained in Sect. \ref{sec:results}, our statistical method does not get confused by these apparent features.

%-----------------------------------------------------------------

%-----------------------------------------------------------------

%***********************
\section{Fitting channel maps} \label{sec:discminer}

\subsection{The Discminer package} 

In this section we present the basic functionality of the \discminer{} code, carefully designed to capture structural and kinematical features from circumstellar discs using molecular line emission. The open source version of the \discminer{} will be released with the second paper of the Disc Miner series. 

\begin{figure*}
   \centering
   \includegraphics[width=0.96\textwidth]{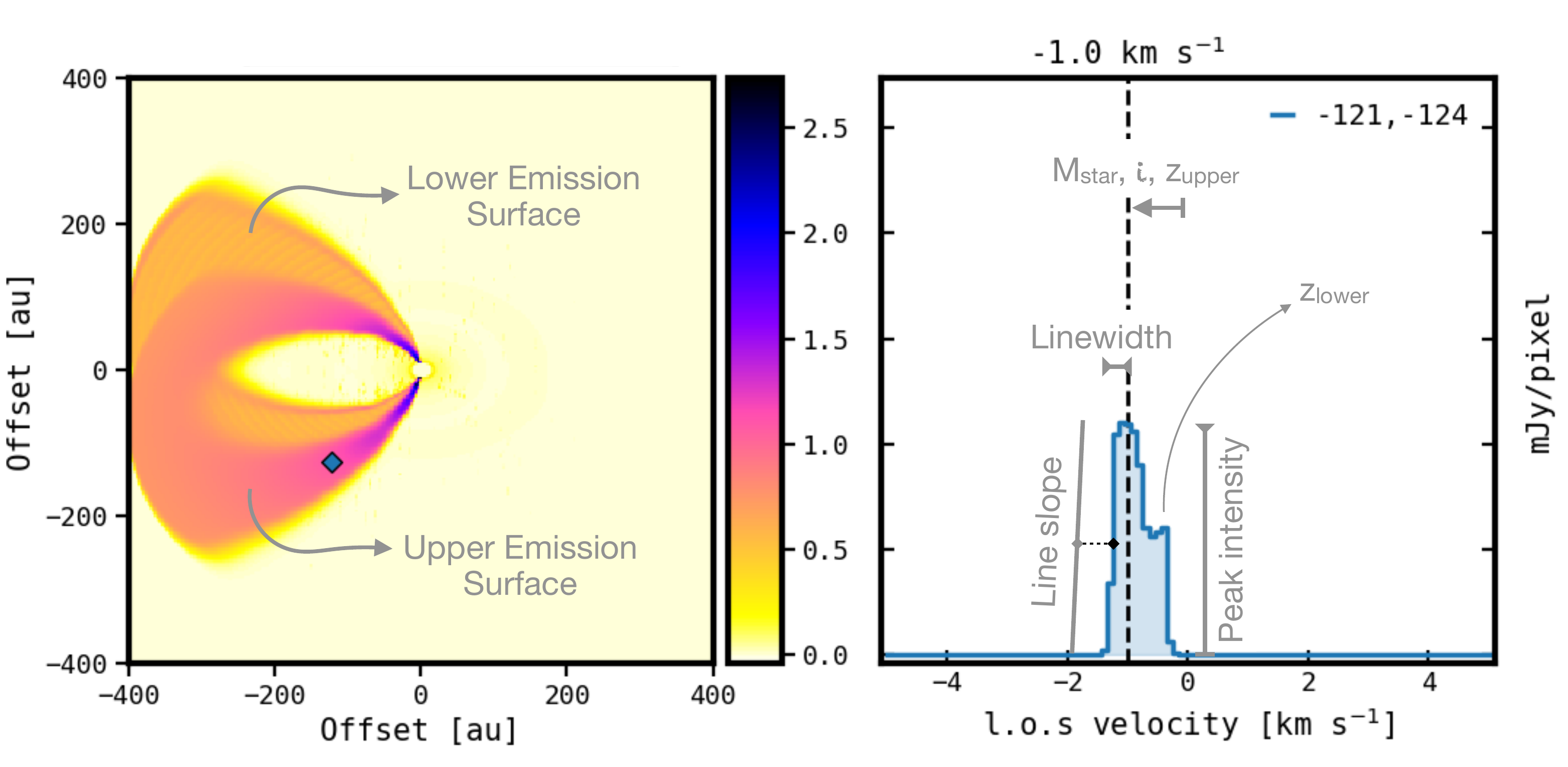}
      \caption{Summary of the main attributes making up the line emission of a disc in the \discminer{}. The left panel shows the projected intensity of the disc for a channel centred on $\upsilon_{\rm ch}=-1.0$\,km\,s$^{-1}$. The right panel is the line intensity profile extracted from the marker on the left. The grey annotations indicate the role of each attribute listed in Table \ref{table:attributes_parameters}. 
              }
         \label{fig:sketch_channel}
   \end{figure*}

As in previous fitting methods that dig into the kinematics of discs \citep[e.g.][]{teague+2018b, casassus+2019}, our package is based on parametric descriptions of the physical and geometrical properties that make up a simplified disc of gas\footnote{The model is designed to be computationally cheap so that it is suitable for parameter space exploration and for analysis of large datasets, while still possessing a reasonable degree of realism.}. However, the \discminer{}  provides for the first time a simultaneous representation of line profiles and velocities by fitting individual intensity channel maps rather than projected velocity maps (see Fig. \ref{fig:sketch_channel}). To do so, the line intensity is modelled considering the following attributes, 
\begin{enumerate}[label=(\roman*)]
    \item We assume that the disc emission comes from two thin (upper and lower) surfaces. Their vertical location is described by two parametric prescriptions of the form $z=f(R)$.
    \item A Keplerian velocity field $\upsilon_{\rm k} = f(R, z; M_\star, \upsilon_{\rm sys})$, which is used to determine the shift in velocity space of the emission from any given pixel. 
    \item A kernel to shape the line intensity profile as a function of the disc coordinates. The kernel combines peak intensity ($I_p$), line-width ($L_w$) and line-slope ($L_s$) information, which are also parametric prescriptions of the form $A = f(R, z)$. See further details in Sect. \ref{subsec:lineprofile}.
    \item Inclination and position angle of the disc. These are used for geometrical projection of intensities and line-of-sight velocities on the plane of the sky.  
\end{enumerate}

The model is then coupled with a Markov chain Monte Carlo (MCMC) random sampler, \textsc{emcee} \citep{foreman+2013}, which efficiently walks over a vast range of parameters to determine the subset of them that best reproduces the projected line intensity of the input disc, often encoded in a three-dimensional ppv cube. We configure the MCMC sampler to maximise a $\chi^{2}$ log-likelihood defined as, 
\begin{equation} \label{eq:chi2}
 \chi^{2} = -0.5\sum_{j}^{n_{\rm ch}} \sum_{i}^{n_{\rm pix}}  w_{i}^{-2} \left[I_m(r_i, \upsilon_j) - I_d(r_i, \upsilon_j)\right]^2,    
\end{equation}
where the index $i$ runs over the pixel location in ($x$, $y$)$_{\rm sky}$ coordinates, and the index $j$ runs over the input velocity channels. The $\chi^{2}$ kernel is the difference between the model intensity, $I_m$, and the input intensity, $I_d$, whose uncertainty is encompassed in the weighting factor $w$ which is computed using residual intensity from line-free velocity channels. 

\subsection{Line intensity profile} \label{subsec:lineprofile}

In this work, we assume a generalised bell function for the line profile kernel, 
\begin{equation}
    I_m(R, z; \upsilon_{\rm ch}) = I_p \left(1+\left|\frac{\upsilon_{\rm ch} - \upsilon_{\rm k^{l.o.s}}}{L_w} \right|^{2L_s} \right)^{-1},
\end{equation}
where $I_p$ is the peak intensity, $\upsilon_{\rm ch}$ is the channel velocity and $\upsilon_{\rm k^{l.o.s}}$ is the Keplerian line-of-sight velocity. The choice of this kernel is motivated by the fact that at the cost of only one additional parameter ($L_s$), it performs better than a Gaussian function at reproducing optically thick lines (such as those from \twCO{} transitions) which are flat at the top and decay rapidly towards the wings. The line width ($L_w$) of the generalised bell function is the half width of the profile at half power. The additional (dimensionless) parameter, the line slope ($L_s$), controls how steep the signal drops at the wings and in turn also determines the spectral extent of the plateau at the top of the profile. It is easy to deduce from the first derivative of the bell function that the slope of a line tangent to either points of the (normalised) profile, at half power, is the ratio $2L_s/L_w$.

For simplicity, we parametrise the peak intensity ($I_p$) and line width ($L_w$) as power laws of the disc cylindrical coordinates ($R$, $z$), which reproduce the overall line profiles of the synthetic observations  reasonably well (see Sect. \ref{sec:results}). The line slope ($L_s$) is allowed to vary but it is assumed uniform everywhere. We note that the model attributes perform well at describing the observed emission on the upper and lower surfaces of the input disc only, and hence any extrapolation to other scale heights should be done with caution.

\subsection{Best-fit model of the simulation}

To reveal deviations from Keplerian rotation, we use the \discminer{} to fit Keplerian channel maps on the simulated synthetic observations. In particular, we obtain three best-fit models, one for each planet mass (0.3, 1.0 and 3.0\,$\Mj$) at $0^\circ$ azimuth. 
To achieve this, we first use the prototyping tool\footnote{This is an interactive tool which allows the user to compare in real time the input data against the model channel maps given a set of attributes and parameters.} of the \discminer{} to find a reasonable set of seeding parameters for \textsc{emcee}. We then let 256 walkers evolve over 1000 steps in a first trial, and allow them to fully converge to the final parameters over another 1500 steps. As our simulations are noise-free, the weighting factor $w$ in Eq. \ref{eq:chi2} is equal to one everywhere. Each run takes about six hours on a 48-core machine with a clock rate of 2.3\,GHz per core. 

\begin{figure}
   \centering
   \includegraphics[width=1.0\columnwidth]{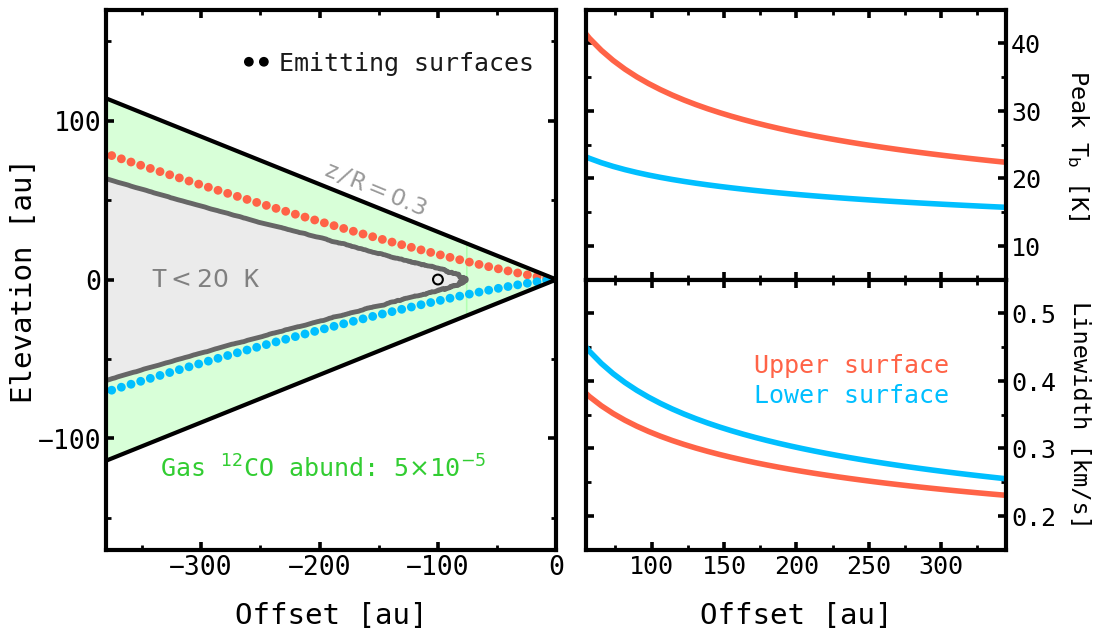}
     \caption{Best-fit attributes obtained by the \discminer{} for the 0.3\,$\Mj$ snapshot. \textit{Left panel:} Freeze-out and gas phase regions for \twCO{}. The coloured circles indicate the height of the model emitting surfaces. \textit{Right panels:} Model peak brightness temperature and line width as a function of radius on both emitting surfaces.
     }
        \label{fig:emission_surfaces}
\end{figure}

Table \ref{table:attributes_parameters} presents a summary of the functional forms considered for the model attributes and the best-fit parameters obtained for the three snapshots. Figure \ref{fig:emission_surfaces} shows the best-fit attributes computed for the 0.3\,$\Mj$ snapshot, deprojected on the upper and lower emitting surfaces of the disc, and the vertical location of both the surfaces (see parameter walkers in Appendix \ref{sec:appendix_figures}). 
We notice slight variations in the best-fit parameters of the other two (1.0 and 3.0\,$\Mj)$ snapshots due to the combined effect of the planet mass and the depth of the gap it carves. More specifically, the stellar mass ($M_\star$) retrieved by the model increases with planet mass ($M_\star=1.013, 1.025$\,M$_\odot$, respectively), while the emission surfaces are shallower because of the increasingly deeper gaps (e.g. $z_0=15.59, 14.35\,{\rm au}$ for the 1.0 and 3.0\,$\Mj$ upper surfaces). Nevertheless, these variations have negligible impact on the detection and quantification of planet perturbations.

Taken together, the parameters retrieved by the model converge notably well to the features we knew beforehand from the input simulations. The mass of the star (1\,M$_\odot$), the disc inclination ($-45^\circ$), and the height of the lower surface tracing the back side of the CO freeze-out region are all closely reproduced even when a planet is present. This means that the \discminer{} is also well suited  to future studies of the three-dimensional structure of discs using molecular line emission.

\setlength{\tabcolsep}{10.5pt} %pad between columns

\begin{table*}
\centering
{\renewcommand{\arraystretch}{1.5}%pad between rows
 \caption{List of attributes considered by the \discminer{} to fit channel maps, and best-fit parameters obtained for the (0.3, 1.0 and 3.0\,$\Mj$, $\phip{}=0^\circ$) synthetic observations.}
  \label{table:attributes_parameters}
\begin{tabular}{ lllll } 

\toprule
\toprule
Attribute & Prescription &  \multicolumn{3}{c}{Best-fit parameters for \twCOfull{}} \\
\midrule

Inclination & $i = i_0$ & $i_0=$ \makecell[l]{$-45.01$ \\ $-44.70$\,deg \\ $-44.41$} & -- & -- \vspace{0.25cm} \\

Rotation velocity & $\upsilon_k = \sqrt{\frac{GM_\star}{R}}$ & $M_\star =$ \makecell[l]{$1.004$ \\ $1.013$\,M$_\odot$ \\ $1.025$} & -- & -- \\ \midrule

Upper surface & $z_U = z_0 (R/100\textrm{\,au})^p$ & $z_0 =$ \makecell[l]{$15.83$ \\ $15.59$\,au \\ $14.35$} & $p=$ \makecell[l]{$1.21$ \\ $1.20$ \\ $1.24$} & -- \vspace{0.25cm} \\

Lower surface & $z_L = z_0 (R/100\textrm{\,au})^p$ & $z_0 =$ \makecell[l]{$13.50$ \\ $13.33$\,au \\ $12.25$} & $p=$ \makecell[l]{$1.24$ \\ $1.24$ \\ $1.28$} & -- \\

\midrule
Peak intensity & $I_p = I_0 (R/100\textrm{\,au})^p (z/100\textrm{\,au})^q$ & $I_0 =$ \makecell[l]{$452.38$ \\ $382.06$\,mJy\,pix$^{-1}$ \\ $442.01$} & $p=$ \makecell[l]{$-4.17$ \\ $-4.02$ \\ $-4.06$} & $q=$ \makecell[l]{$3.19$ \\ $3.10$ \\ $3.10$} \vspace{0.25cm} \\

Line width & $L_w = L_{w0} (R/100\textrm{\,au})^p (z/100\textrm{\,au})^q$ & $L_{w0} =$ \makecell[l]{$61.4$ \\ $65.4$\,m\,s$^{-1}$ \\ $69.0$} & $p=$ \makecell[l]{$0.81$ \\ $0.77$ \\ $0.74$} & $q=$ \makecell[l]{$-0.90$ \\ $-0.86$ \\ $-0.82$} \vspace{0.25cm} \\ 

Line slope & $L_s = L_{s0}$ & $L_{s0} =$ \makecell[l]{$4.21$ \\ $4.12$ \\ $4.35$} & -- & -- \\

\bottomrule

\end{tabular}

  }
\end{table*}

%-----------------------------------------------------------------
%-----------------------------------------------------------------

%***********************
\section{Finding the kinematical footprints of planets} \label{sec:results}

\subsection{Intrinsic deviations from Keplerian rotation}

%-----------------------------------------------------------------
\begin{figure}
   \includegraphics[width=0.97\columnwidth]{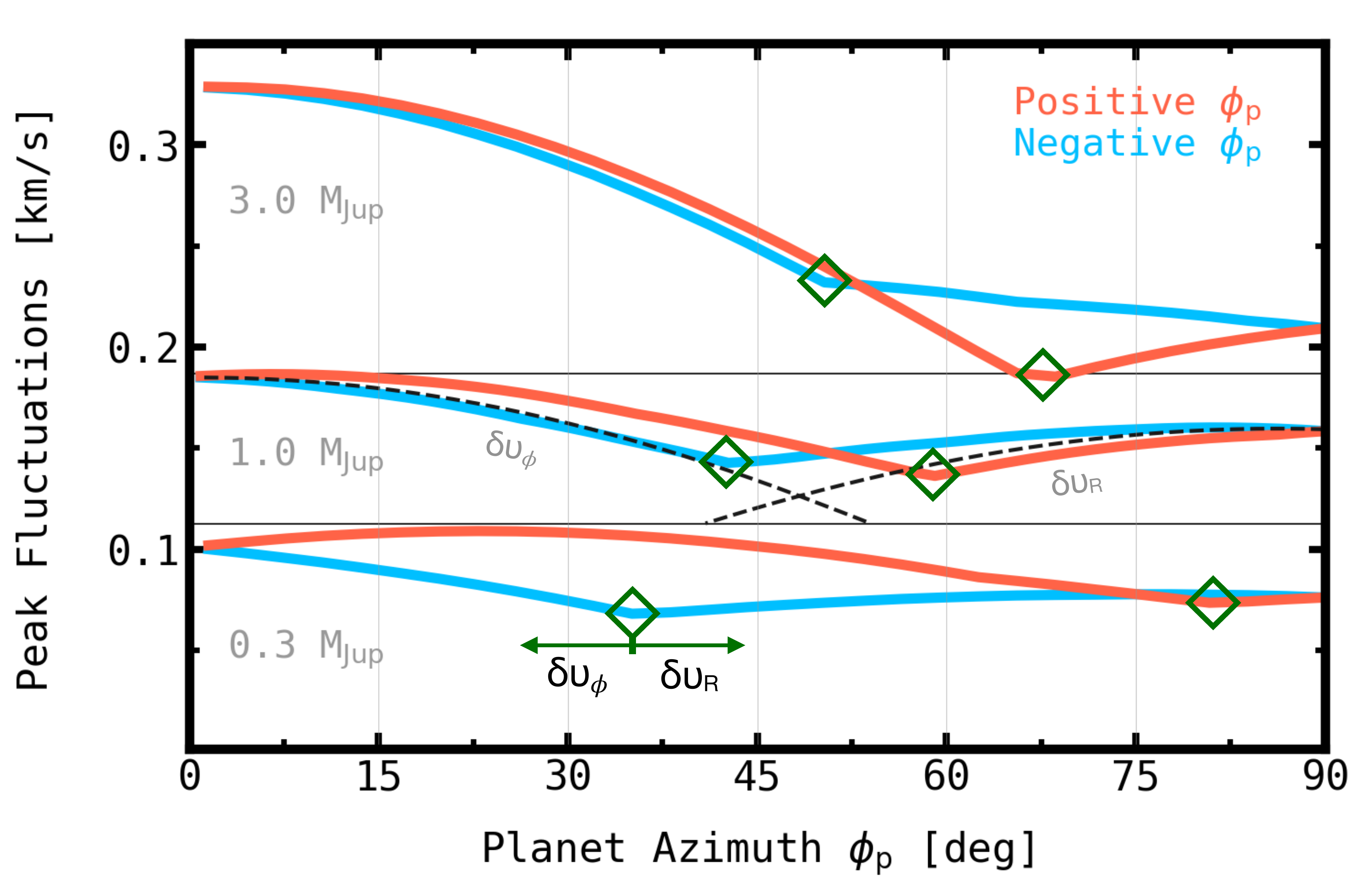} 
     \caption{Peak line-of-sight deviations from Keplerian rotation versus positive and negative planet azimuths, for all three planet masses, extracted from the native simulations (without radiative transfer effects). The diamonds highlight turnover points where the radial velocity perturbations become important and eventually dominant over the projected azimuthal perturbations.
              }
         \label{fig:native_fluctuations}
\end{figure} 
%-----------------------------------------------------------------

Before analysing observables, it is worth looking at the pristine simulation velocities to understand the posterior effect of radiative transfer and disc structure on the retrieved planet-driven perturbations. In Figure \ref{fig:native_fluctuations}, we present peak line-of-sight velocity deviations from Keplerian rotation (in magnitude) for all three planet masses, varying planet azimuths from $-90^\circ$ to $90^\circ$. We use the same disc inclination ($i=-45^\circ$) adopted for the synthetic observations, and assume a line-of-sight parallel to the vertical axis. There are two overall features that stand out here. First, it is clear that the projected peak deviations are different for positive and negative planet azimuths. This is because the outer and inner spiral wakes launched from the planet, are geometrically different, and aside from direction, perturbations along these spirals generally differ in magnitude as well. Hence one or the other dominates depending on the planet mass (inner[outer] spiral perturbations are higher for the 0.3[1.0, 3.0]\,$\Mj$ planet), and on the projected direction of the fluctuations at each planet azimuth. Second, in all cases there is a sharp turnover azimuth where the radial component of the perturbation starts dominating over the azimuthal one. This is explained by the fact that, unlike azimuthal perturbations, the highest radial perturbations do not necessarily occur along spiral wakes but on circumplanetary material (see Fig. \ref{fig:dens+temp}, right panel), implying that the radial and azimuthal perturbations generally contribute independently to the projected perturbation.

In the following sections, we perform a similar analysis on synthetic observations of the simulations and find that radiative transfer and disc structure can strongly influence the observed velocity fluctuations as compared to the actual gas velocities.

\subsection{Residual maps}

\begin{figure*}
\begin{center}
\begin{tabular}{l}

 \includegraphics[width=1.0\textwidth]{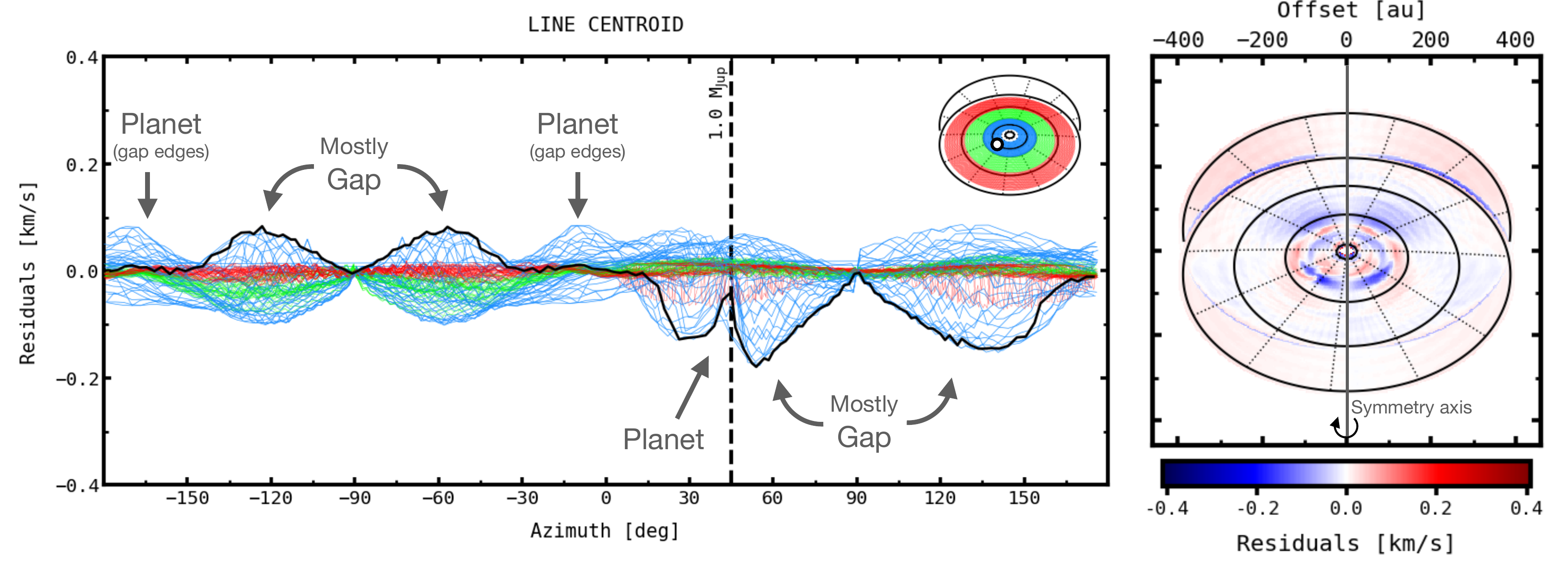} \\
 \includegraphics[width=1.0\textwidth]{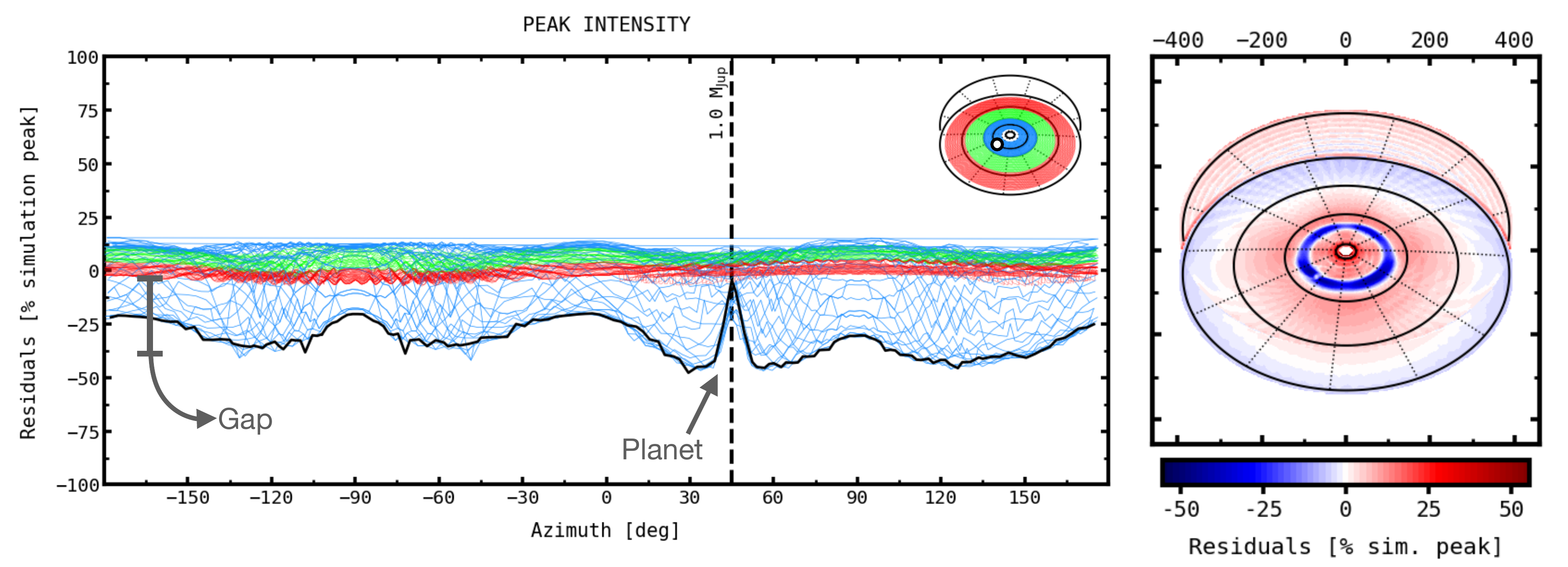} \\
 \includegraphics[width=1.0\textwidth]{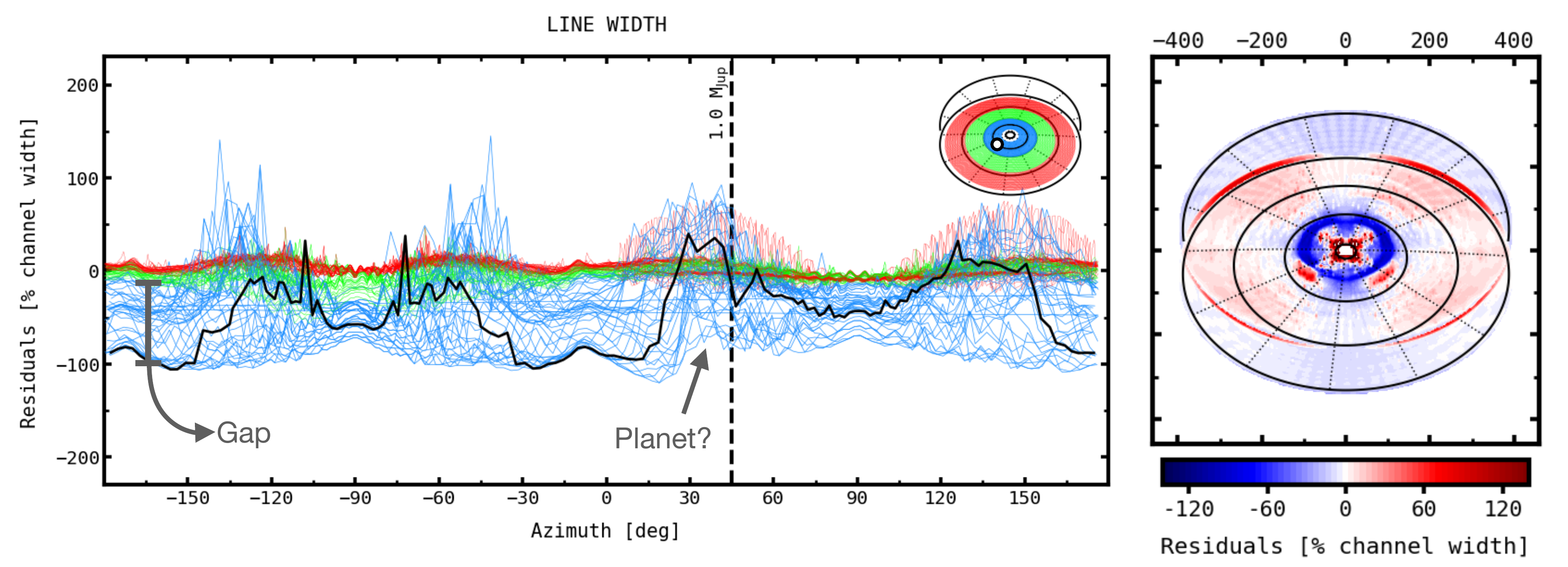} 
 
\end{tabular}
\caption{Line centroid (top), peak intensity (middle), and line width (bottom) residuals for the (1.0\,M$_{\rm Jup}$, $\phip{}=45^\circ$) snapshot. The azimuthal scans on the left run along constant radii contours in disc coordinates; their colours represent their closeness to the radial location of the planet, with blue being the closest. The solid black contour runs along the projected distance of the planet ($R_p=100$\,au) and the dashed black line shows the azimuth of the planet.
} 

\label{fig:azimuth_residuals}
\end{center}
\end{figure*}

In order to extract observables and quantify any line profile differences between the simulation and the smooth Keplerian (best-fit) model, we fit a Gaussian profile to each pixel on both the simulation and model intensity cubes for each planet mass and azimuth (see Appendix \ref{sec:appendix_gap}). The mean, amplitude, and standard deviation of each Gaussian profile are assumed to be the line centroid, peak intensity, and line width of the corresponding pixel. 
We then subtract the best-fit model line profile properties from those of the simulations and produce residual maps as illustrated in Figure \ref{fig:azimuth_residuals}. 
There, we conveniently present a full scan of residuals from one of our snapshots as a function of azimuth along constant radii contours whose colours represent their closeness to the actual radial distance of the planet ($R_p=100$\,au). 

%-----------------------------------------------------------------
%-----------------------------------------------------------------

A number of features  of these residual maps are worthy of discussion. First, the three types of residuals are all mostly uniform along the green and red contours which correspond to the outermost parts of the disc, meaning that the simulation and the model line profiles are close to identical in regions away from the planet. Second, residuals along the blue contours are rather high and are mainly associated with the gap carved by the planet (see Appendix \ref{sec:appendix_gap}), whose contribution is not considered by the smooth Keplerian  model. However, in all cases   these residuals are symmetric along the projected minor axis (hereafter the `symmetry axis') of the disc, $\phi=\pm 90^{\circ}$. Unsurprisingly, this symmetry is notably disrupted by the contribution of the embedded planet, whose strongest effect is localised both in radius and azimuth. In Sect. \ref{subsec:centroid_folding}, we exploit these (a)symmetries to isolate kinematical perturbations driven by planets as a function of their mass and azimuth in the disc, and distinguish them from other perturbations.

\subsection{Line centroid folding} \label{subsec:centroid_folding}

Taking advantage of the fact that the strongest kinematical perturbations driven by planets are spatially and spectrally localised, we propose a line centroid folding method to remove any symmetric contribution to the velocity field arising from the natural rotation of the disc and from the large-scale contribution of the gap. The method consists of subtracting line centroid velocities from one half of the disc from those of the other half, exactly as if the disc was folded along its symmetry axis, $\phi=\pm90^\circ$. 

\begin{figure*}
\includegraphics[width=0.36\textwidth]{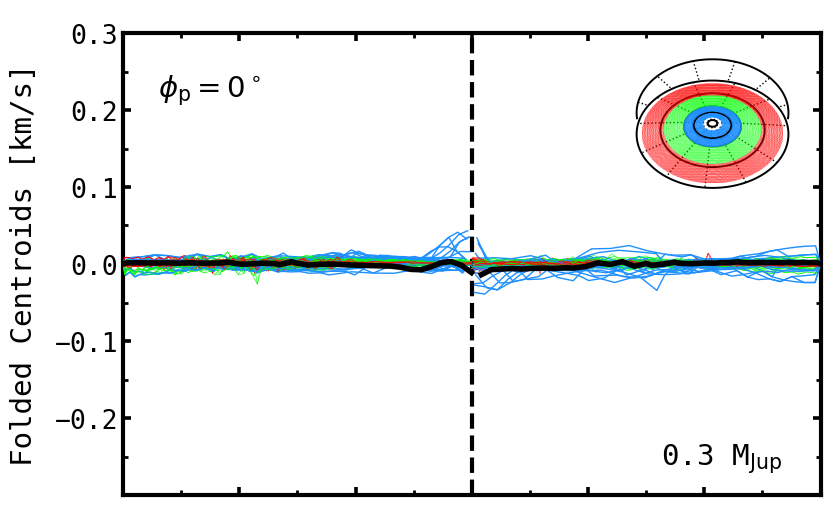}
 \hspace{0.02cm}
\includegraphics[width=0.312\textwidth]{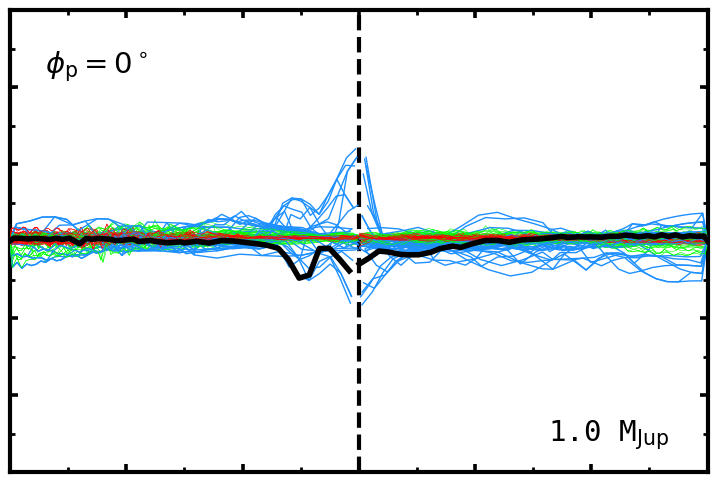}
 \hspace{0.02cm}
\includegraphics[width=0.312\textwidth]{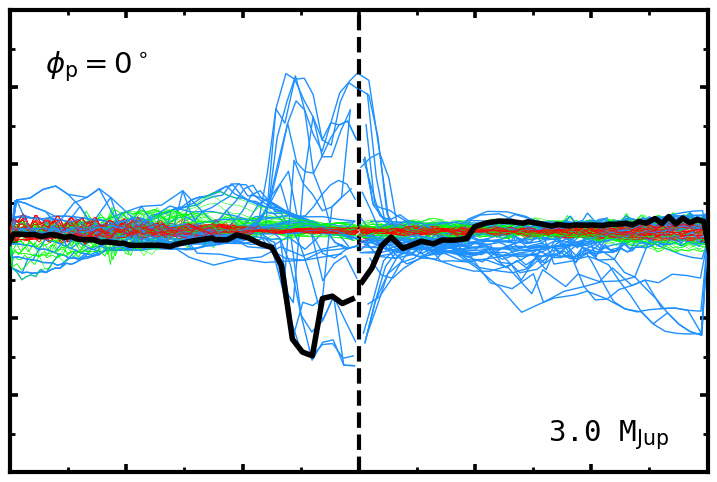}

\includegraphics[width=0.36\textwidth]{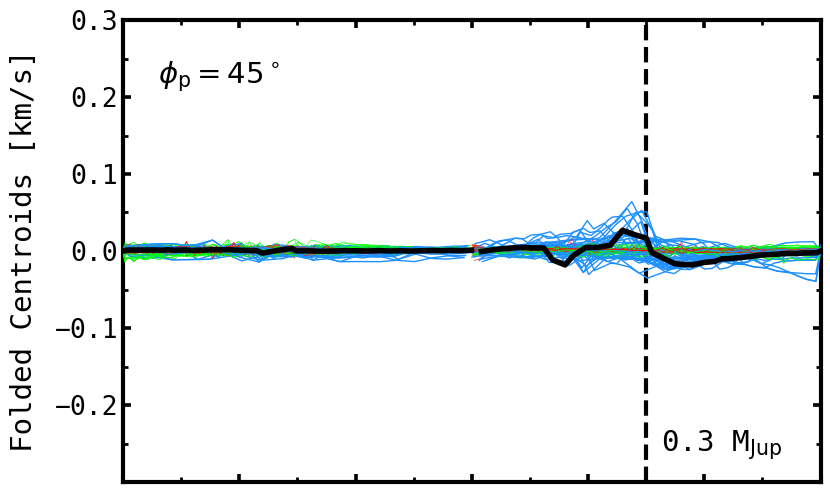}
 \hspace{0.02cm}
\includegraphics[width=0.312\textwidth]{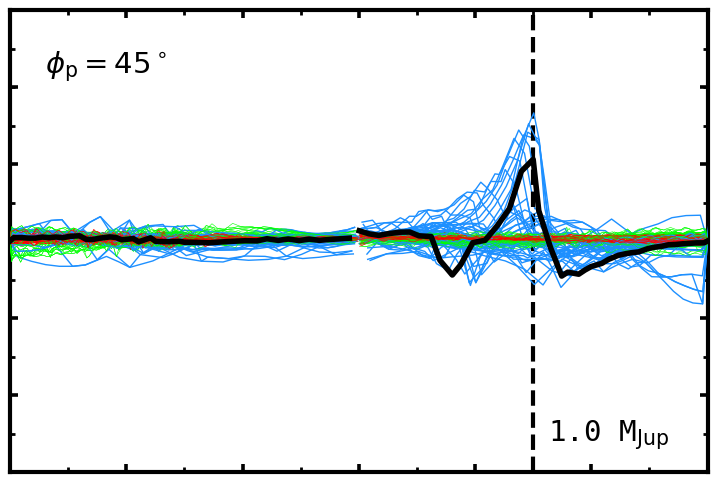}
 \hspace{0.02cm}
\includegraphics[width=0.312\textwidth]{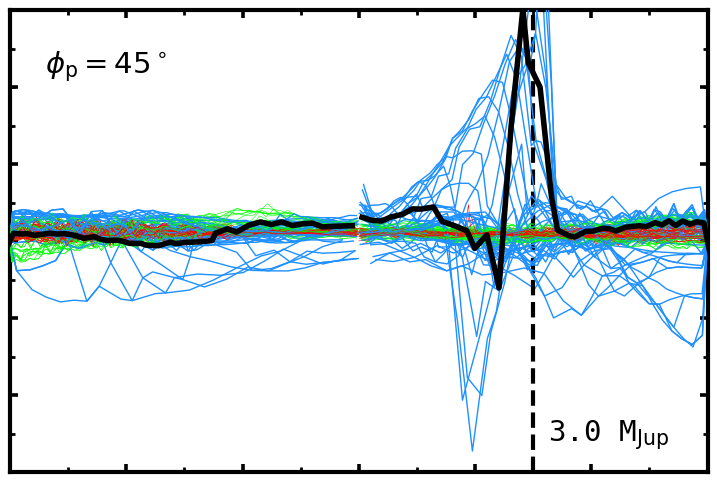}
 \includegraphics[width=0.367\textwidth]{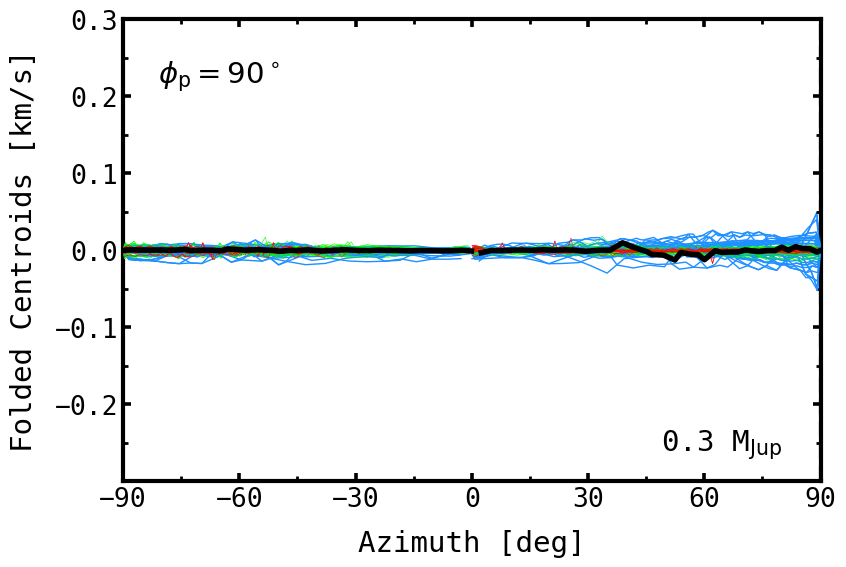}
 \hspace{-0.1cm}
\includegraphics[width=0.318\textwidth]{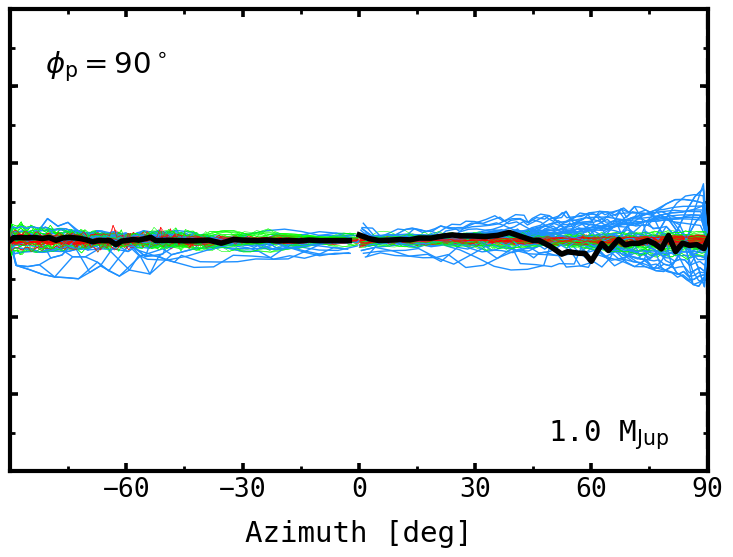}
 \hspace{-0.1cm}
\includegraphics[width=0.318\textwidth]{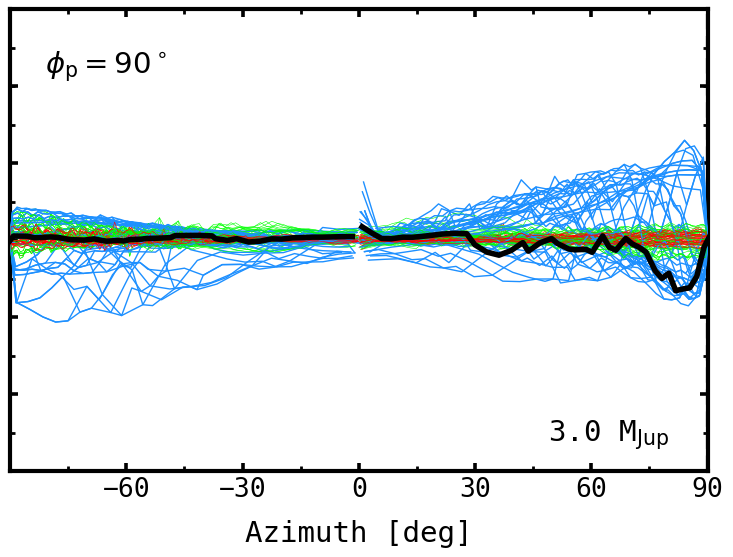}
\caption{Folded (magnitude of) line centroid residuals as a function of azimuth, varying the analysis radius (in colours), for different planet masses (0.3, 1.0 and 3.0\,$\Mj$, from left to right) and azimuths (0$^\circ$, 45$^\circ$, and 90$^\circ$ from top to bottom). The solid black contour runs along the projected distance of the planet ($R_p=100$\,au), and the dashed black line corresponds to the azimuth of the planet.} 

\label{fig:azimuthal_folded_centroids}
\end{figure*}

We illustrate the outcome of folding line centroids in Figure \ref{fig:azimuthal_folded_centroids}, for all three planet masses, namely 0.3, 1.0 and 3.0\,$\Mj$, and three azimuthal locations, 0$^\circ$, 45$^\circ$ and 90$^\circ$. Unlike in the raw residual maps, the highest velocity residuals are this time closely related to the embedded planet thanks to the localised nature of the perturbation, whereas most of the contribution from the gap is cancelled out. Of particular interest, the magnitude of the residuals, as well as their spatial extent, appear to be tightly linked to the mass and azimuthal location of the planet. Quantitatively, typical peak centroid residuals range within 40$-$70\,m\,s$^{-1}$, 70$-$170\,m\,s$^{-1}$, and 130$-$450\,m\,s$^{-1}$ for the 0.3, 1.0, and 3.0\,$\Mj$ planets, respectively. At $\phip{}=90^\circ$ azimuth, all planets trigger the lowest velocity residuals, and from $\phip{}=30^\circ$ to $60^\circ$, the highest. The angular dependence of residuals can be readily understood by noting that at $\phip{}=90^\circ$ and neighbouring angles, most of the azimuthal component of the planet perturbation is cancelled out because it is orthogonal to the line-of-sight.  
The equivalent occurs around $\phip{}=0^\circ$ azimuth for the radial component of the perturbation.
Nevertheless, interpreting the high-velocity fluctuations observed at intermediate angles is less straightforward. The observed fluctuations at these angles do not only depend on the intrinsic magnitude of the perturbation, but also on the scale height at which the perturbation is measured and hence on the structure of the disc itself. This effect is further discussed in Section \ref{sec:discussion_magnitude_pert}. 

%-----------------------------------------------------------------
%-----------------------------------------------------------------

Now, we discuss whether or not it is possible to determine not only the presence, but also the actual location of the planets from the information gathered immediately above.
From Fig. \ref{fig:azimuthal_folded_centroids}, this appears to be the case, especially for those planets at azimuths smaller than 90$^\circ$, where the perturbation is obvious to the eye. However, line centroid residuals from less massive planets, such as the 0.3\,$\Mj$ included in our analysis, may be particularly challenging to detect just by visual inspection because the localised signature, intrinsic to the planet, is small and can easily get confused with other asymmetric, though more extended perturbations. Also, planets of any mass near the symmetry axis of the disc, $\phi=90^\circ$, are equally challenging because their signature is no longer seen as localised to the naked eye. In the following section, we provide two statistical methods that conveniently examine line centroid residuals and do not rely on visual inspection to robustly infer the location of embedded planets in discs.

\subsection{Statistical methods to detect planets} \label{subsec:statistical_detection} 
In order to detect an embedded planet through kinematical analysis, one needs to be able to spatially isolate the velocity fluctuations that the planet induces on the gas disc, but also to provide a robust measurement of their magnitude. Here we present two statistical methods to detect and quantify planet-driven kinematical perturbations using synthetic observations of simulations with different planet masses and azimuths (see Sect. \ref{sec:simulations_rt}). Both methods make use of the folded centroid residuals presented in Sect. \ref{subsec:centroid_folding}. 

From inspection of Fig. \ref{fig:azimuthal_folded_centroids}, a first natural approach would be to determine the location of the global maximum centroid residual, which is apparently well correlated with the azimuthal and radial location of the planet. Following this idea, we extract the peak velocity residual for each radius in the disc as a function of the azimuth where it occurs. From the resulting distribution of velocity residuals, we compute a $3\sigma$ threshold below which residuals are discarded from the analysis. 
The inferred location of the planet is assumed to be the median azimuth and the median radius of the leftover peak residuals, as illustrated in Figure \ref{fig:spot_planet_1Mjup_PA45}. This procedure is what we call the Global Peak detection method. In Appendix \ref{fig:fitmass}, we show the dependence of peak fluctuations as functions of planet mass by fitting $\delta\upsilon=aM^b$ powerlaws, for each planet azimuth, with $a$ and $b$ as free parameters. We note that there are substantial differences between the observed fluctuations at negative and positive planet azimuths. As explained later in Sect. \ref{sec:discussion_magnitude_pert}, the reason is that the disc vertical structure and the gap ---combined with projection effects--- are important, and their contribution to the observed velocity fluctuations depends on the location of the planet.

%-----------------------------------------------------------------
\begin{figure*}
   \centering
   \includegraphics[width=1.0\textwidth]{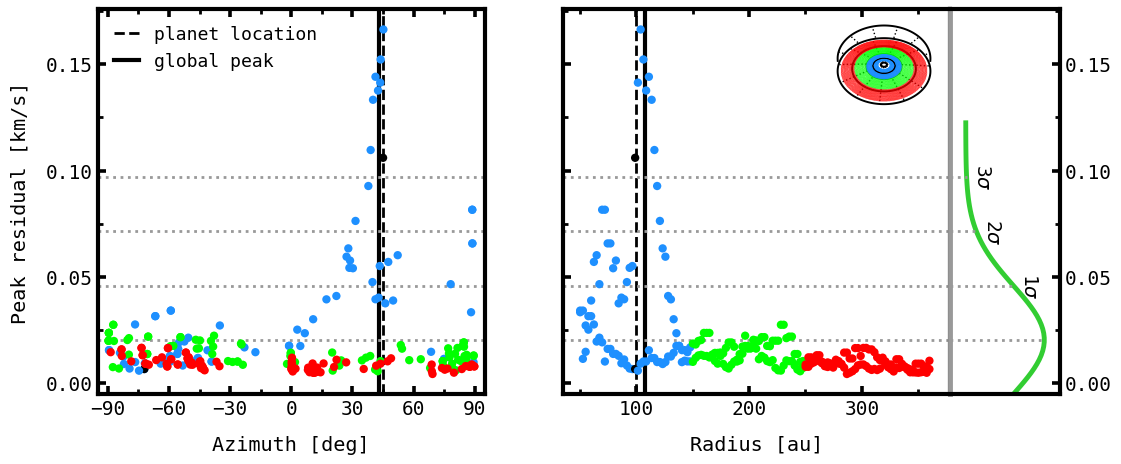}
      \caption{\textit{Left panel}: Peak residuals and their azimuthal location for the (1.0\,M$_{\rm Jup}$, $\phip{}=45^\circ$) snapshot. \textit{Right panel}: Peak residuals rearranged as a function of their radial distance. The panel on the right shows the peak residual distribution (green line) and 1, 2, and 3\,$\sigma$ significance thresholds (dotted lines). The solid black lines are the azimuthal and radial location of the global peak, while the dashed black lines show the actual location of the planet.
              }
         \label{fig:spot_planet_1Mjup_PA45}
\end{figure*} 
%-----------------------------------------------------------------

\begin{figure*}
   \centering
   \includegraphics[width=1\textwidth]{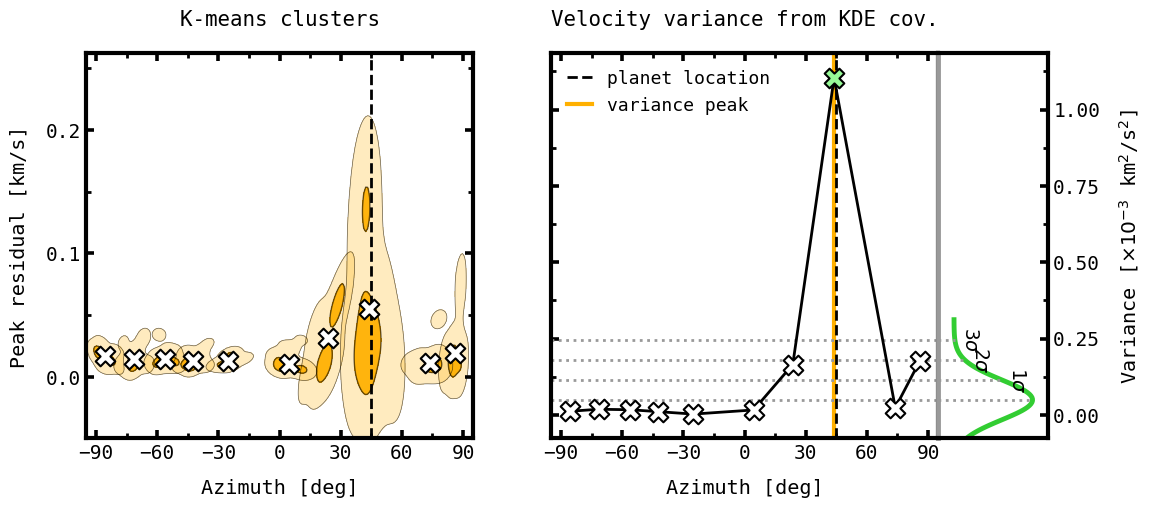}
   \includegraphics[width=0.28\textwidth]{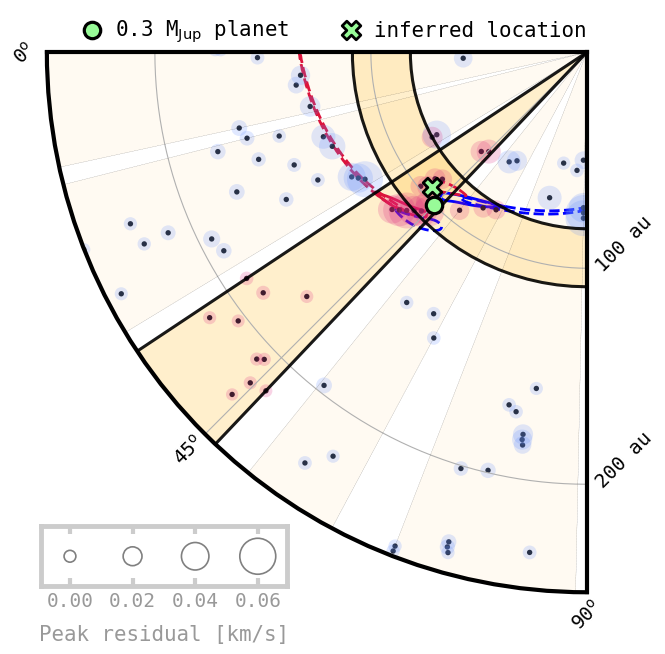}
   \includegraphics[width=0.4\textwidth]{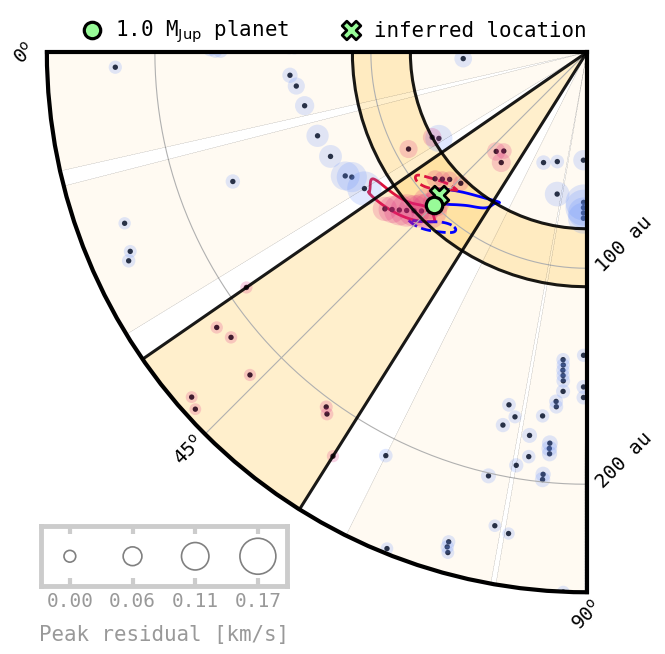}
   \includegraphics[width=0.28\textwidth]{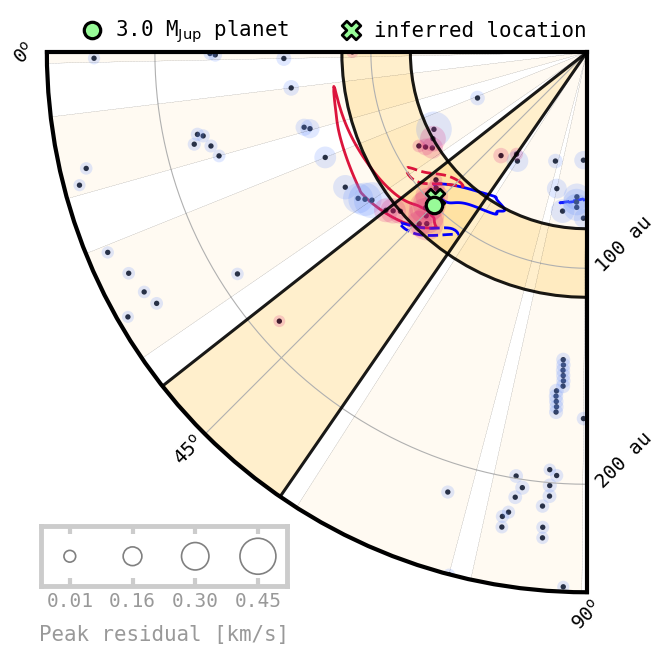}
      \caption{Top row. \textit{Left panel}: KDE contours derived from K-means clusters of peak velocity residuals for the (1.0\,M$_{\rm Jup}$, $\phip{}=45^\circ$) snapshot. The contours enclose 33 and 95 per cent of the peak residuals. The white crosses are the cluster centres retrieved by the K-means algorithm.  
      \textit{Right panel}: Spectral variance of peak residuals for each cluster on the left. In the panel attached on the right we show 1, 2, and 3$\sigma$ significance thresholds. The yellow line highlights the azimuthal location of the peak variance and the dashed line the actual location of the planet. The green cross is the peak variance significant enough to be attributed to the planet perturbation. Bottom row: 2D visualisation of the detection using the Variance Peak method for three planets (0.3, 1.0 and 3.0\,$\Mj$), all at $\phip{}=45^\circ$ azimuth and $R_p=100$\,au radius. The green circles and crosses are the actual and inferred location of the planets, respectively. The highlighted sectors are the azimuthal and radial clusters with the highest spectral variance extracted by the method. The boundaries of the sectors mark the maximum spatial coverage of the clusters. The circles show the location of the observed peak velocity residuals, whose magnitude is indicated by their size. The red circles are residuals within the azimuthal or radial peak variance clusters. The solid ($\delta\upsilon_\phi$) and dashed ($\upsilon_R$) contours trace spiral wakes and radial planet perturbations, and correspond to 60 per cent peak velocity fluctuations extracted from the simulation velocities. 
              }
         \label{fig:clusters_1Mjup_PA45}
\end{figure*}

Typically, after comparing the inferred and the actual location of planets, the Global Peak method is accurate within $\pm 3^\circ$ in azimuth and $\pm 8$\,au in radius, but its significance, which in all cases is between $3\sigma$ and $7\sigma$, might be a weak spot in noisy scenarios. This leads us to introduce a rather different method, based on the assumption that the strongest velocity fluctuations driven by planets should be simultaneously localised and coherent, which in other words implies that the velocity field, as well as its observables, should vary smoothly as one approaches the planet. If this is true, in addition to large velocity residuals there should also be a higher density of peak velocity residuals around the planet compared to undisturbed regions of the disc. This is pictured in Figure \ref{fig:clusters_1Mjup_PA45}, where we find clusters of peak residuals, independently for each spatial coordinate ($R$, $\phi$), using a K-means clustering algorithm \citep{macqueen+1967, lloyd+1982}. The K-means algorithm subdivides the input residuals into a predefined number of clusters in such a way that the centre of each cluster is the closest centre to all the residuals in the cluster. In other words, the input data are iteratively partitioned into Voronoi cells until convergence is reached, which in this case means until the sum of squared distances from the data to the centre of their clusters is minimised. We note that the minimisation distance used by the iterative procedure to identify clusters in the top panel of Fig. \ref{fig:clusters_1Mjup_PA45} is defined in 2D (azimuth, velocity residual) space, but in practice the azimuthal distance dominates over the distance in velocity residuals so that the clustering is similar to a binning in azimuth (the same applies for the radial clustering). However, unlike simple manual binning, the bin boundaries are in this case irregular, and the bin centres tend to be near the densest accumulation of points, which is ideal for localising coherent velocities. 
We adopt ten clusters such that the azimuthal and radial extent of the localised perturbation from the planets ($\delta\phip{}\approx\!20^\circ\!-\!50^\circ$, $\delta R\approx\!50-100$\, au, at a radius of 100\,au) is always within one to three clusters. 
Next, the cluster with the highest velocity variance is attributed to the planet-driven perturbation as long as it meets the requirement of being above the standard 3$\sigma$ threshold, which refers to the variances of the background clusters. In this case, the retrieved azimuthal location of the planet is the azimuth of the centre of the cluster with the peak variance. The same recipe is followed to infer the radial location of the planet\footnote{We note from Fig. \ref{fig:clusters_1Mjup_PA45} that, around planets, peak velocity residuals trace planet-driven spiral wakes and part of the circumplanetary material.}. 
This method, the Variance Peak, is equally accurate to the Global Peak method when it comes to determining the location of planets. However, it strikingly boosts the significance of the detection thanks to the coherent nature of velocities around planets.

%-----------------------------------------------------------------

\begin{figure*}
   \centering
   \hspace{-0.14cm}
   \includegraphics[width=1.0\textwidth]{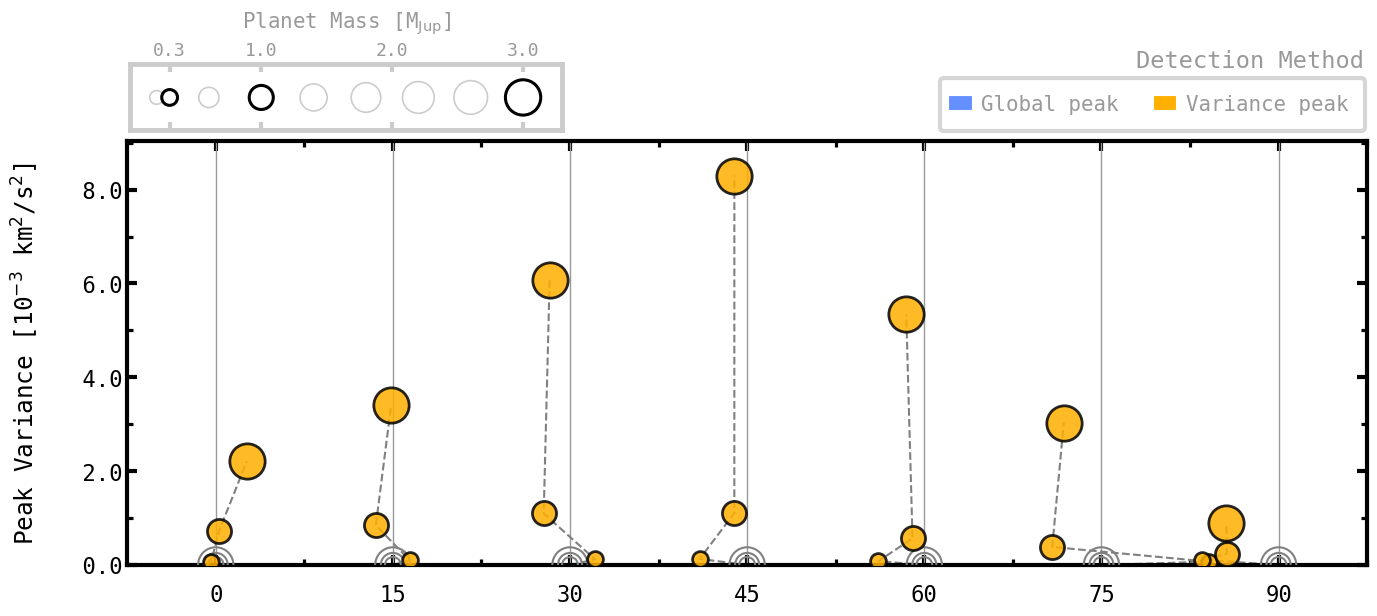} \\
   \includegraphics[width=0.995\textwidth]{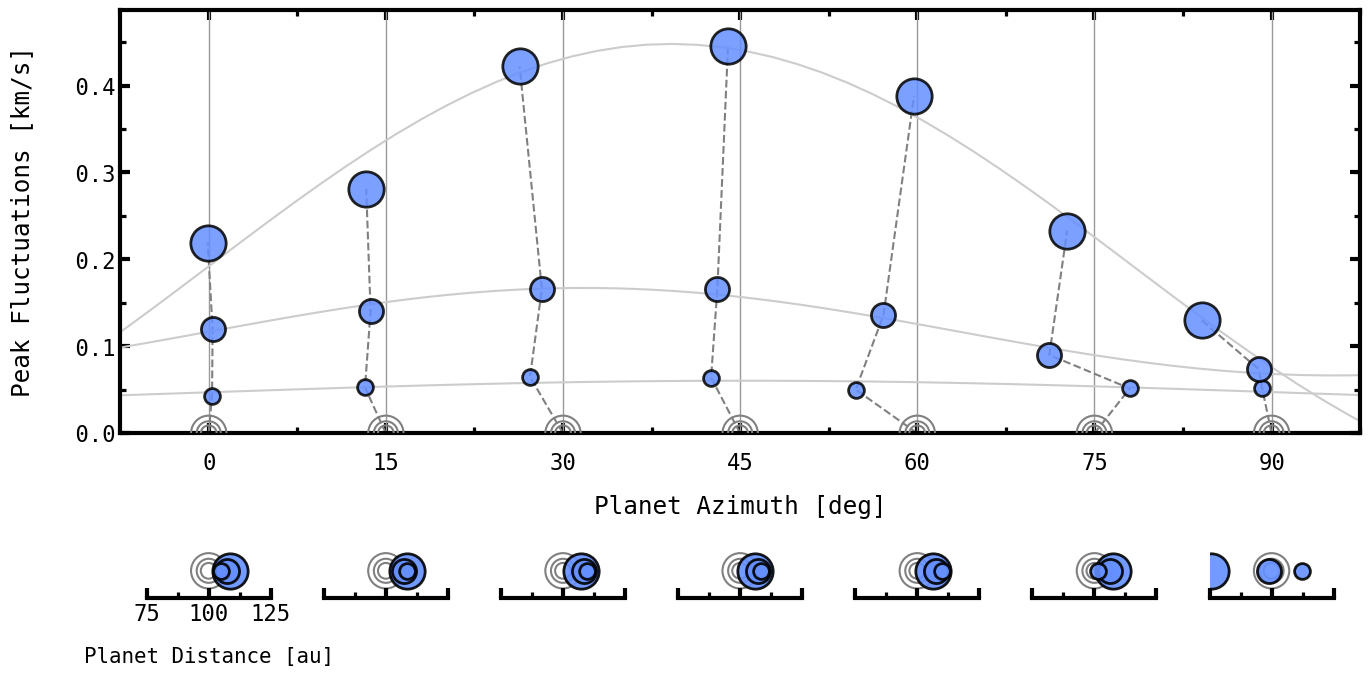}
     \caption{Peak variance and peak velocity fluctuations extracted with the Variance and Global peak methods as a function of the detected planet azimuth for all three planet masses. The detected radial distance is shown in the bottom panels. Empty circles indicate the actual locations of planets.
              }
         \label{fig:detection_summary_globalpeakvalue}
\end{figure*}

The absolute value of peak residuals in the Global Peak method, and the peak variances in the Variance Peak method are presented in Figure \ref{fig:detection_summary_globalpeakvalue}. Unsurprisingly, regardless of the planet azimuth, peak fluctuations and peak variances increase steadily with the mass of the planet. Also, because of projection effects, the highest peaks are reached at intermediate angles (30$^\circ$, 45$^\circ$, 60$^\circ$). 
Furthermore, all of the detected radial locations are approximately within $R=100\pm10$\,au except for the $\phip{}=90^\circ$ snapshots where there is a higher dispersion of peak residuals, making it more challenging for the technique to find the radial location of the planet. Interestingly, we note that for the same $\phi=90^\circ$ snapshots, the azimuthal location of all planets is reasonably well determined, as opposed to the first impression left by the bottom row of Fig. \ref{fig:azimuthal_folded_centroids}, where velocity residuals were anything but localised to the naked eye. 

\begin{figure*}
\begin{center}

 \includegraphics[width=1.0\textwidth]{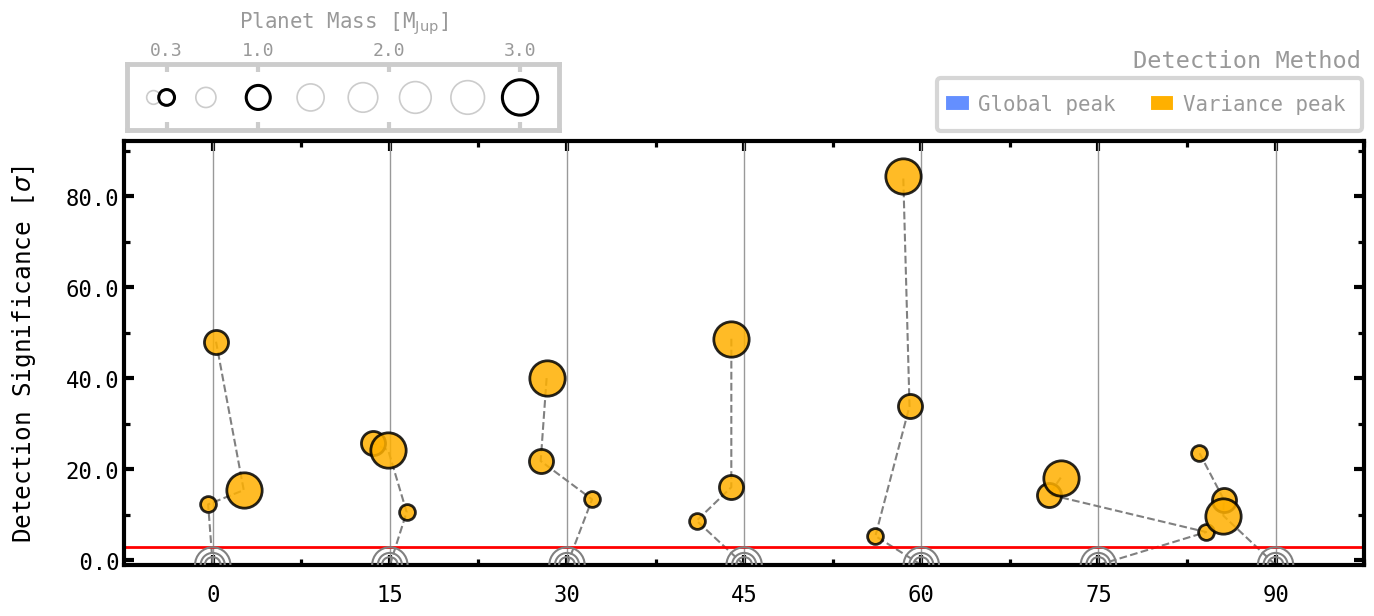}
 \includegraphics[width=1.0\textwidth]{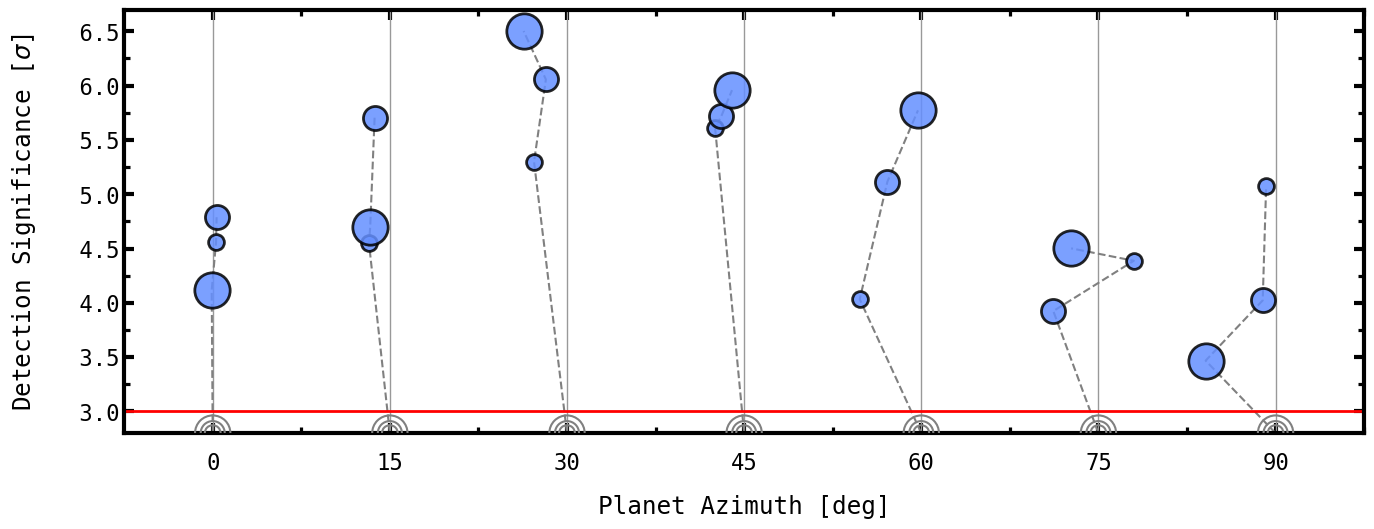}
\caption{Significance of planet detections using the Variance and Global peak methods as a function of planet azimuth for all three planet masses. 
The red line highlights the 3$\sigma$ threshold considered to validate a detection. Empty circles indicate the actual location of planets.} 

\label{fig:detection_summary}
\end{center}
\end{figure*}

However, of greater interest is the statistical significance of both methods. Even though both approaches work well at inferring the location of planets for all masses and azimuths, the Variance Peak is far more robust than the Global Peak method. As illustrated in Figure \ref{fig:detection_summary}, the significance of the Variance Peak detections are in almost all cases larger (or even much larger) than $10\,\sigma$, whereas the Global Peak detections are always between $3\,\sigma$  and  $7\,\sigma$. Again, this is thanks to the coherence of the velocity field around planets leading to a significant accumulation of peak residuals to which the Variance Peak method is fairly sensitive. The Global Peak method, on the other hand, is not sensitive to the bulk of velocity residuals around planets, but is limited to only the highest of them.

We note that massive planets such as the 3.0\,$\Mj$ planet in our sample do not always yield the highest detection significance. The reason behind this is mainly that the velocity perturbation from massive planets is so spatially extended that some of the velocity residuals associated with the planet are actually contributing to the background residuals, which in turn reduces the significance of the detected global and variance peaks. We softened this effect in the Variance Peak method by allowing more than one cluster to be considered as part of the planet perturbation so that the background is better constrained prior to calculation of the significance of the detection. In any case, the fact that the least massive planets in our sample (0.3, 1.0\,$\Mj$) are in almost all cases robustly detected is encouraging, given that such low masses are almost undetectable by empirical methods. 

%-----------------------------------------------------------------
%-----------------------------------------------------------------

%-----------------------------------------------------------------

%***********************
\section{Discussion} \label{sec:discussion}

\subsection{Comparison with other methods}

The fitting methods encompassed in the \discminer{} have proven to be effective at describing the large-scale structure of discs but also to be accurate when detecting localised signatures on intensity and velocity driven by embedded planets. In particular, our fitting technique has a number of advantages over previous modelling efforts. 
First, it can describe intensity, line width, and velocity field simultaneously. Second, it is able to reproduce the height and line properties of both the upper and lower emitting surfaces of the disc independently, which is essential to properly assessing the flux and kinematics of discs. Third, it consistently gains velocity accuracy by fitting multiple channel maps at once, allowing for subspectral measurements of small-scale perturbations even on cubes with standard spectral resolutions (e.g. $\Delta_{\rm chan}=100$\,m\,s$^{-1}$).

Modelling all these ingredients together is key to understanding the actual contribution from embedded planets to the velocity perturbations, whose observables are secondary products that depend on the underlying intensity of the disc. To illustrate this, in Appendix \ref{sec:appendix_gap} we manually removed the planet from one of our simulations and imposed the gas velocities to be either fully Keplerian or non-Keplerian but azimuthally symmetric. We then noticed that the wavy behaviour of the velocity residuals displayed in Fig. \ref{fig:azimuth_residuals} is mostly driven by the gap, or more specifically, by the fact that the differences between simulation and model line intensities on the upper and lower emitting surfaces of the disc are more prominent at the location of the gap. This discrepancy causes the resulting line centroids to be red- or blueshifted in relation to one other depending on their projected location only. One way to work around this is by comparing velocities at peak intensities (namely on the upper surface of the disc only), but at the cost of a substantial loss of velocity accuracy as this method is closely dependent on the channel width of the data. 
Additionally, as explained in Sect. \ref{sec:discussion_magnitude_pert}, the observed velocity perturbations may be stronger and more clearly projected on the lower surface of the disc depending on the planet location, in which case they would go unnoticed by peak intensity methods. Our line centroid folding procedure gets rid of the unwanted contribution of the gap while keeping both (sub)spectral resolution and information from the lower surface of the disc. Equivalently, as illustrated in Appendix \ref{sec:appendix_gap}, this procedure allows us to distinguish between azimuthally localised planet-driven fluctuations and axisymmetric non-planet-induced velocity perturbations, with the second scenario effectively leading to non-detections. 

On the other hand, previous techniques that rely on visual inspection of kink-like features \citep[e.g.][]{pinte+2018b, pinte+2020} are unable to find low-mass planets ($< 1\,\Mj$) or planets of any mass near the projected minor axis of the disc, because kinks are camouflaged with the background velocities in both of those contexts. Our statistical method can instead capture localised, coherent velocities, 
and therefore detect planets even when kinks are not visually manifest. This also implies that even though random noise or large-scale fluctuations may be comparable in magnitude to the localised planet-driven velocity fluctuations, the latter are more densely assembled and therefore more easily detected by our technique. 

\subsection{Robustness of the spectral resolution of the method}

We assessed the robustness of the spectral resolution of our method by performing synthetic observations with the same setup as in Sect. \ref{sec:simulations_rt} but using half the original channel width this time (i.e. $\Delta_{\rm chan}=50$\,m\,s$^{-1}$). The detected planets and the significance of the measurements remained almost the same except for a single snapshot (0.3 $\Mj$, $\phip{}=75^\circ$), where the detection was indeed closer to the actual location of the planet. This suggests that our analysis is already very close to measuring velocity fluctuations without suffering from channelisation effects. 

\subsection{Magnitude of the observed planet-driven perturbations} \label{sec:discussion_magnitude_pert}

Our findings suggest that the observable velocity fluctuations driven by planets depend on the intrinsic magnitude of the perturbations, given by Fig. \ref{fig:native_fluctuations}, but also on the disc vertical structure and the depth of the gap carved by the planet.

\begin{figure*}
   \includegraphics[width=1\textwidth]{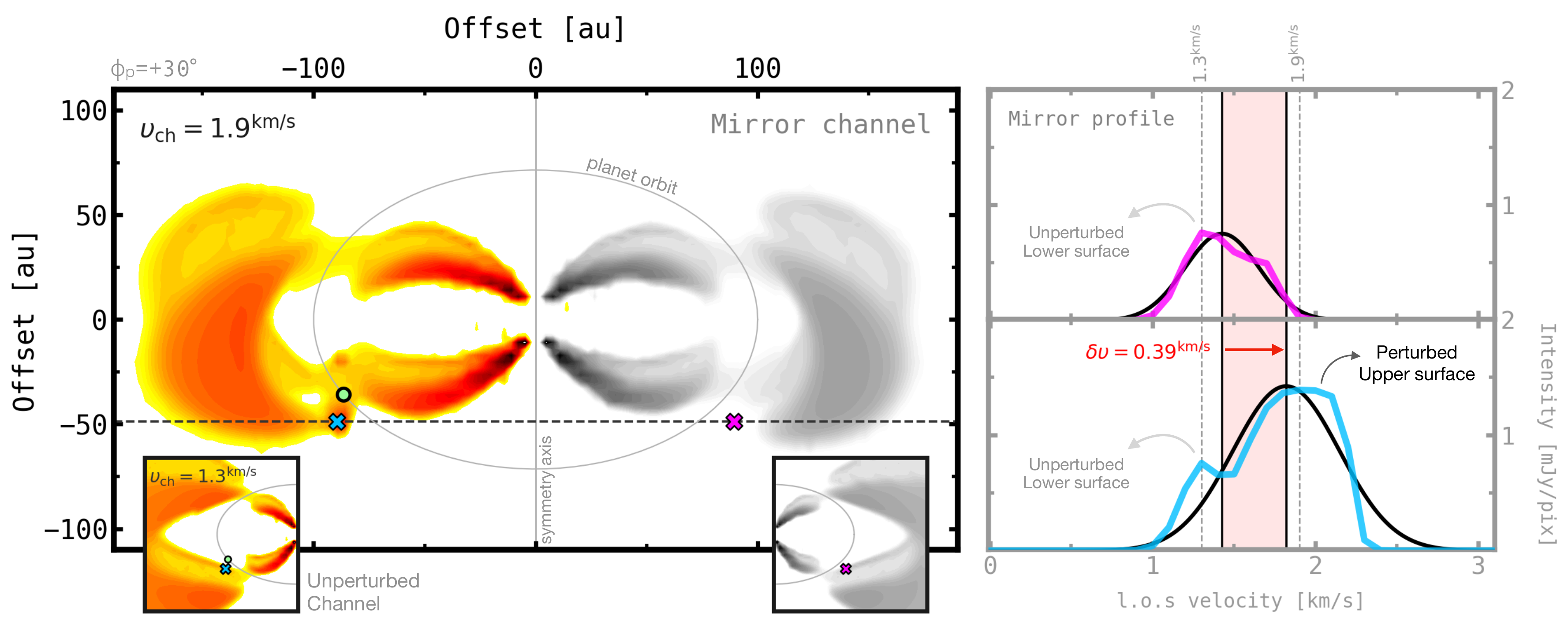}
   \includegraphics[width=1\textwidth]{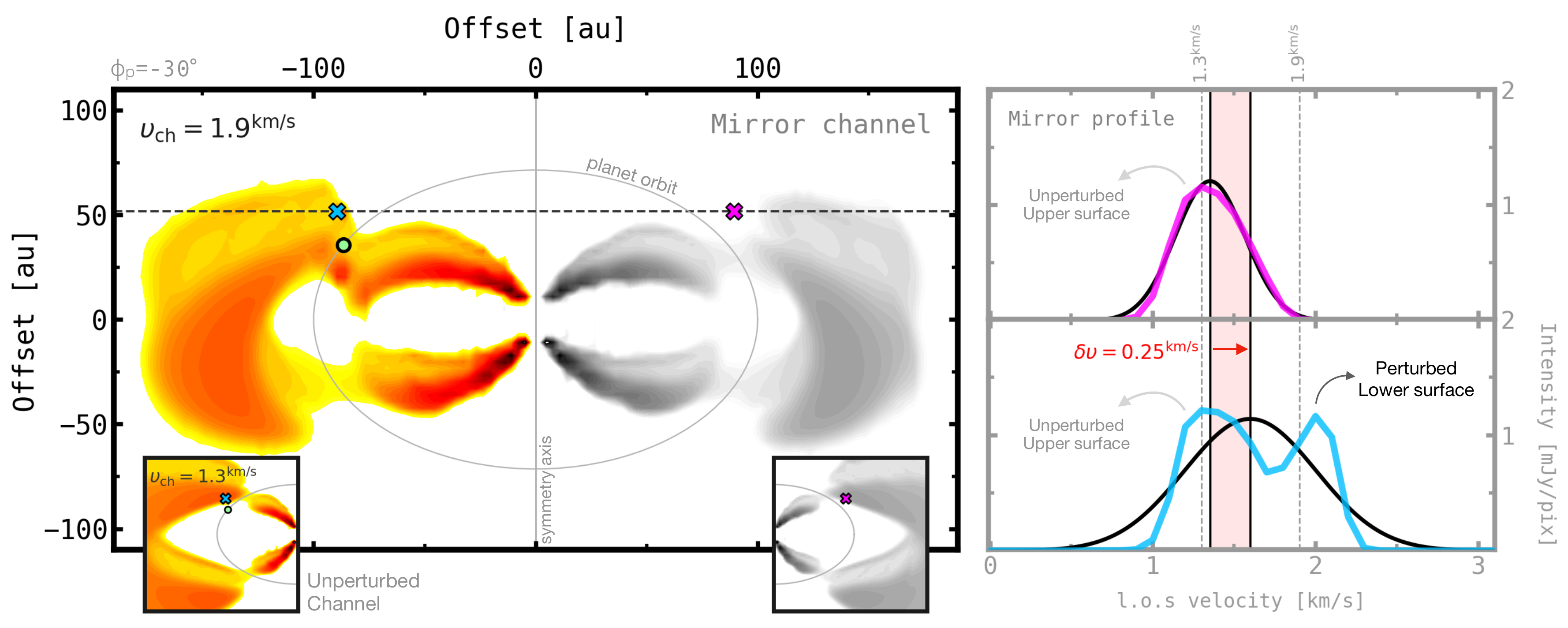}
   
     \caption{Illustrating how line centroids are shifted towards faster velocities around a 3.0\,$\Mj$ planet, at two azimuths $\phip{}=\pm30^\circ$. The crosses on the left indicate the location of the line profiles with the same colours on the right. The blue cross is the location of the observed peak perturbation, and the pink cross is the mirror pixel at the other side of the disc. We note that depending on the planet azimuth, the peak perturbation is projected either on the upper or on the lower emitting surface, which contribute differently to the observed line profiles. This contrast leads to variations in the observed line centroids (vertical black lines on the right) and the retrieved velocity fluctuations (in red).
              }
         \label{fig:channel_pos_neg}
\end{figure*}

Figure \ref{fig:channel_pos_neg} demonstrates the impact of disc structure on the observed velocity fluctuations for a $3.0\,\Mj$ planet at $\phip{}=+30^\circ$ and $\phip{}=-30^\circ$ azimuths. Despite the fact that the {intrinsic} peak fluctuations are almost the same for both angles, the {observed} peak fluctuations differ significantly. To understand this, we first highlight the fact that the emission height of the perturbation determines the background velocity (at the other side of the disc) against which it will be compared to compute the magnitude of the perturbation. If the test background velocity falls within the gap, it will trace a different Keplerian velocity from the one it would trace at the same emitting surface of the perturbation. More specifically, perturbations projected outside the orbit of the planet will be overestimated, whereas those projected inside the orbit will be underestimated, by a factor that depends on the depth of the gap. For a similar reason, the observed peak fluctuations are generally dominated by one of the emitting surfaces of the disc, which at the same time determines how much the line profile centroid can be shifted. 
In particular, the brighter, upper surface of the disc shifts line centroids more than the fainter, lower surface. This effect is especially prominent for massive planets, which carve deeper gaps, and for intermediate planet azimuths, where the projected perturbation is more extended and the observed line profiles are clearly shaped by both emitting surfaces. No less important is the role of the disc inclination, which determines how much of the lower surface can be observed and how far apart the emission from both surfaces is in the velocity space. 

%-----------------------------------------------------------------
%-----------------------------------------------------------------

With this example as background information, it is worth discussing some of the trends obtained for the observed peak velocity fluctuations. As illustrated in Fig. \ref{fig:detection_summary_globalpeakvalue}, planets at intermediate angles are better detected and yield the highest velocity residuals, $\delta\upsilon=70,170,450$\,m\,s$^{-1}$ for 0.3, 1.0, and 3.0\,$\Mj$, respectively. The retrieved velocity fluctuations are enhanced in those cases by the combined contribution of the gap and projection effects on the upper and lower emitting surfaces of the disc, as explained above. Overall, the retrievable velocity fluctuations from planets seem periodic as a function of azimuth, and the amplitude of such a pattern correlates well with the mass of the planet. 
It is also interesting to analyse the limiting planet azimuths $\phip{}=0^\circ$, $90^\circ$  because they provide constraints on the orthogonal components of the perturbation by cancelling out one of them by line-of-sight projection. For the $\phip{}=0^\circ$ case, we obtain peak residuals $\delta\upsilon_\phi=40,120,220$\,m\,s$^{-1}$ for each planet mass, where the subscript $\phi$ indicates that the observed perturbation is nearly fully azimuthal. Similarly, for the $\phip{}=90^\circ$ snapshot we find peak residuals $\delta\upsilon_r=50,75,130$\,m\,s$^{-1}$. These limiting cases better match the trend of the intrinsic peak deviations from Keplerian presented in Fig. \ref{fig:native_fluctuations}, where the azimuthal perturbations are often stronger than the radial ones. However, we note that the observed peak perturbations tend to be lower than the intrinsic ones. This is because most of the background reference gas at the opposite side of the perturbation stands on the non-Keplerian edges of the gap, which in turn softens the magnitude of the observed perturbation around the planet. 

\subsection{Caveats}

We warn the reader that we do not claim to present a full characterisation of planet-driven perturbations, nor do we include an exhaustive list of all variables that real discs encompass. Instead, the goal of the paper is to provide a new methodology for detecting embedded planets through careful inspection of line emission profiles under certain disc conditions. Nevertheless, the technique can potentially be extrapolated to other disc scenarios. As such, to keep degeneracy at its lowest level, we did not include a full three-dimensional treatment of the gas velocities. We assumed simple cylindrical rotation implying that the central force from the star, per unit mass, is simply $f_R=GM_\star/R^2$. This means that there is no differential rotation along the vertical coordinate of the disc, which leads to errors of between $\sim\!40-60$\,m\,s$^{-1}$ in the rotation velocities as measured on the model emitting surfaces. However, because of the axisymmetric nature of this effect, it does not have any impact on the detection of planet-driven perturbations with our method. On the other hand, the vertical gravitational pull from the planet and hydrodynamic meridional flows driven by the carved gap are not considered either. Also, some observational biases such as noise and beam smearing are excluded from the analysis. Future releases of the Disc Miner series more focused on observations will incorporate these effects.

%***********************
\section{Conclusions} \label{sec:conclusions}

We introduce a novel statistical technique to detect kinematical perturbations driven by embedded planets in discs. The method is sensitive to localised deviations from Keplerian rotation by examining line centroid differences using intensity channel maps, which allows us to locate and quantify velocity fluctuations around planets with high accuracy, all while preserving line width and intensity information. Our approach is powered by the \discminer{} package, which aims to model channel maps by simultaneously fitting the intensity, rotation velocity, and height of the upper and lower emitting surfaces of the disc. The package was originally developed for kinematical analyses, but it is also well suited to studying the three-dimensional structure of discs. 

We tested this new method on synthetic observations of the \twCOfull{} line from simulations of planet--disc interactions to explore variations in the disc kinematics as a function of the planet mass (0.3, 1.0 and 3.0\,$\Mj$) and planet azimuth (from $-90^\circ$ to $90^\circ$ in steps of $15^\circ$) for a disc inclination of $-45^\circ$. As expected, the observed velocity fluctuations increase with planet mass. 
In all cases, the highest deviations from Keplerian rotation are found at intermediate azimuths, $\phip{}=\pm$(30$^\circ$, 45$^\circ$, 60$^\circ$), which are strongly enhanced by the influence of the gap and the vertical structure of the disc combined with projection effects. 
The lowest velocity fluctuations are obtained for planets on or near the projected main axes of the disc (i.e. along $\phi=0^\circ$ or $\phi=\pm90^\circ$).   
The method can detect all planets at all azimuths, despite the fact that some of them do not exhibit clear kinks in the channel maps, such as the 0.3\,$\Mj$ planet, or planets of any mass near the projected minor axis of the disc, $\phi=\pm90^\circ$.
Likewise, our approach does not get confused by apparent kinks triggered by gaps in the gas disc or, more generally, by any axisymmetric velocity perturbation field, regardless of its origin.

Our technique takes advantage of the coherence of velocity fluctuations around planets to boost the detection of planet-driven kinematical perturbations in gas discs. We find this to be a substantial improvement to the previous methods for determining whether or not localised velocity fluctuations are unambiguously manufactured by embedded planets.

\begin{acknowledgements}
The authors would like to thank the anonymous referee for their constructive comments that helped improve the manuscript.
This work was partly supported by the Italian Ministero dell Istruzione, Universit\`a e Ricerca through the grant Progetti Premiali 2012 – iALMA (CUP C$52$I$13000140001$), 
by the Deutsche Forschungs-gemeinschaft (DFG, German Research Foundation) - Ref no. FOR $2634$/$1$ TE $1024$/$1$-$1$, 
and by the DFG cluster of excellence Origins (www.origins-cluster.de). 
This project has received funding from the European Union's Horizon 2020 research and innovation programme under the Marie Sklodowska-Curie grant agreement No 823823 (DUSTBUSTERS) and from the European Research Council (ERC) via the ERC Synergy Grant {\em ECOGAL} (grant 855130). SF acknowledges an ESO fellowship. GR acknowledges support from the Netherlands Organisation for Scientific Research (NWO, program number 016.Veni.192.233) and from an STFC Ernest Rutherford Fellowship (grant number ST/T003855/1). This work was performed using the DiRAC Data Intensive service at Leicester, operated by the University of Leicester IT Services, which forms part of the STFC DiRAC HPC Facility (www.dirac.ac.uk). The equipment was funded by BEIS capital funding via STFC capital grants ST/K000373/1 and ST/R002363/1 and STFC DiRAC Operations grant ST/R001014/1. DiRAC is part of the National e-Infrastructure.
\end{acknowledgements}

% WARNING
%-------------------------------------------------------------------
% Please note that we have included the references to the file aa.dem in
% order to compile it, but we ask you to:
%
% - use BibTeX with the regular commands:
%   \bibliographystyle{aa} % style aa.bst
%   \bibliography{Yourfile} % your references Yourfile.bib
%
% - join the .bib files when you upload your source files
%-------------------------------------------------------------------
\bibliographystyle{aa}
\bibliography{references}

\begin{appendix}

\section{Impact of the gap and emitting surfaces on the observed gas velocities} \label{sec:appendix_gap}

\begin{figure*}
\begin{center}
\begin{tabular}{l}

 \includegraphics[width=1.0\textwidth]{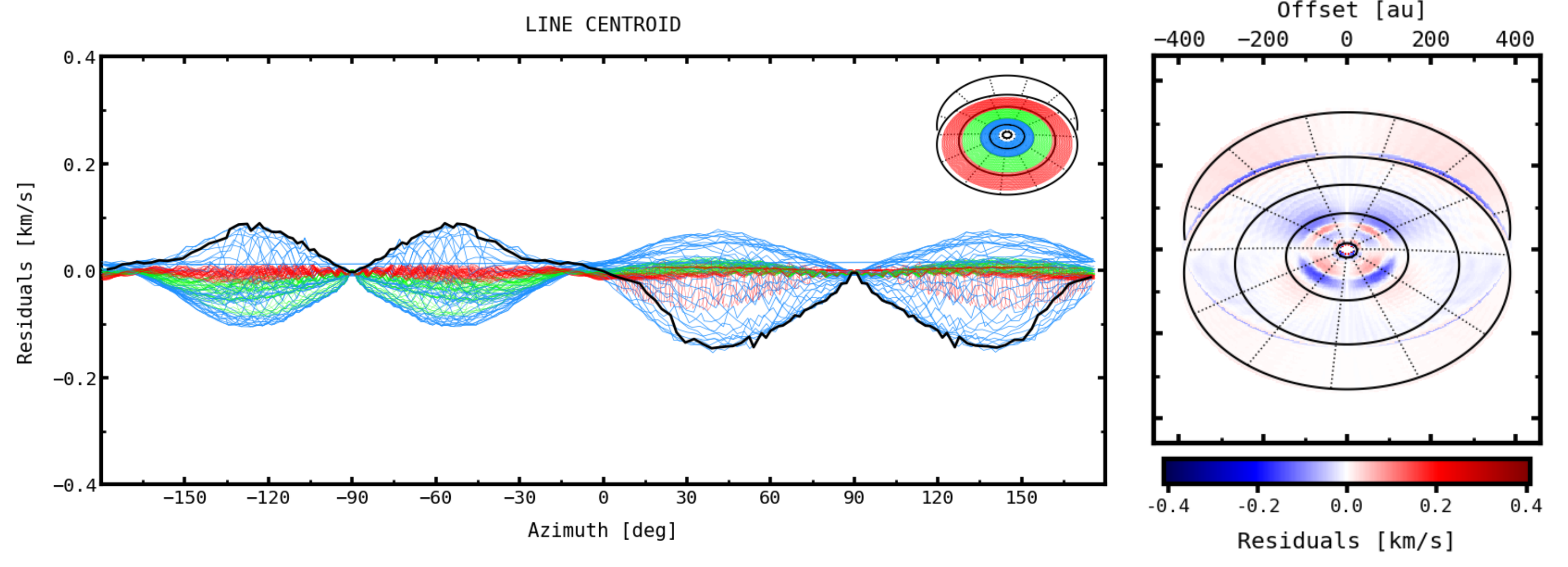} \\
 \includegraphics[width=1.0\textwidth]{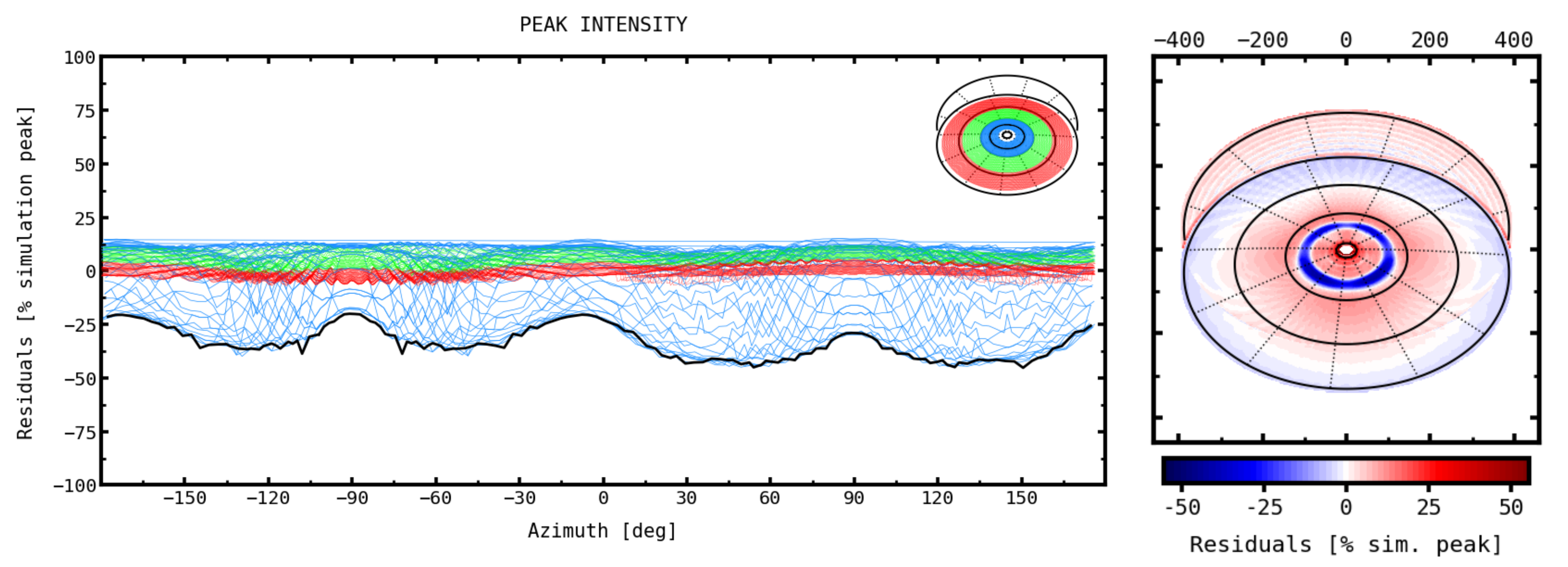} \\
 \includegraphics[width=1.0\textwidth]{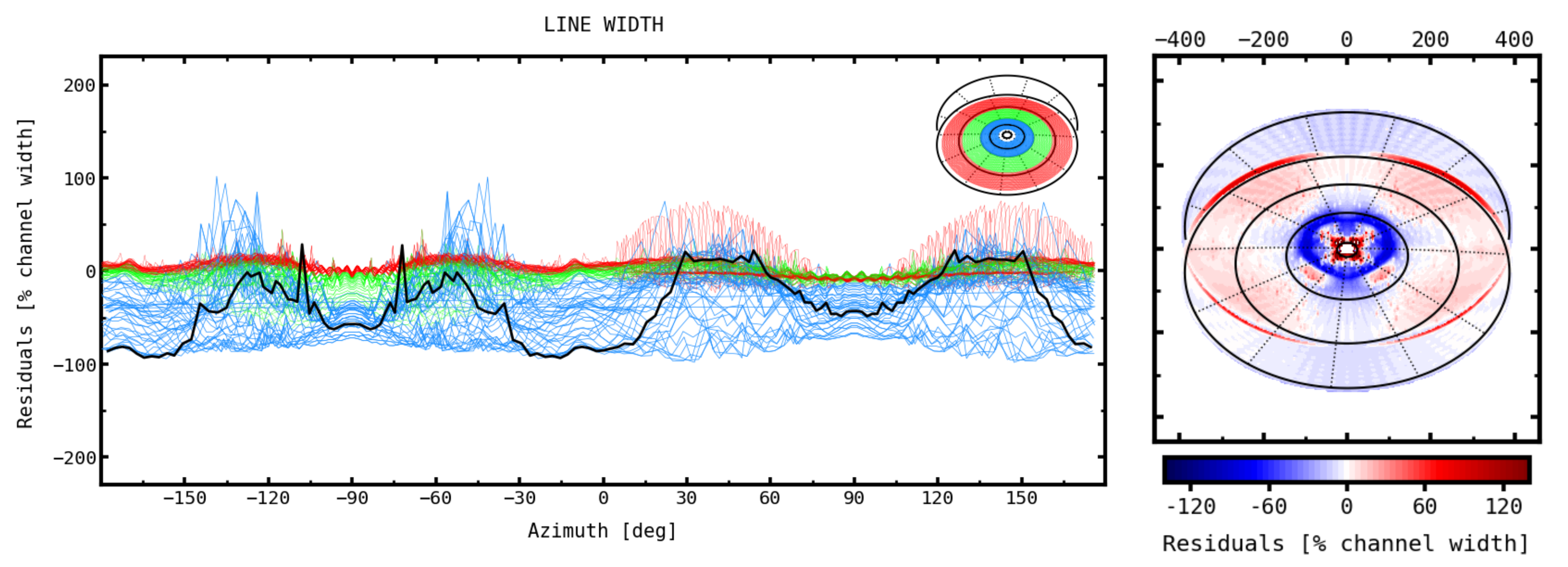} 
 
\end{tabular}
\caption{Line centroid (top), peak intensity (middle), and line width (bottom) residuals for the 1.0\,$\Mj$ snapshot. The planet was removed and the gas velocity set to be fully Keplerian in order to analyse the contribution of the gas gap alone. The azimuthal scans on the left run along constant radii contours, whose colours represent their closeness to the gap, centred at $R=100$\,au. Although the velocities are Keplerian, the high (symmetric) centroid velocity residuals remain due to intensity differences between simulation and model at the location of the gap (see Fig. \ref{fig:gaussian_profiles_gap}). By simple comparison with Fig. \ref{fig:azimuth_residuals}, it is easy to identify the impact of the planet on the intensity and velocity residuals.} 

\label{fig:azimuth_residuals_noplanet}
\end{center}
\end{figure*}

\begin{figure*}
   \centering
   \includegraphics[width=0.84\textwidth]{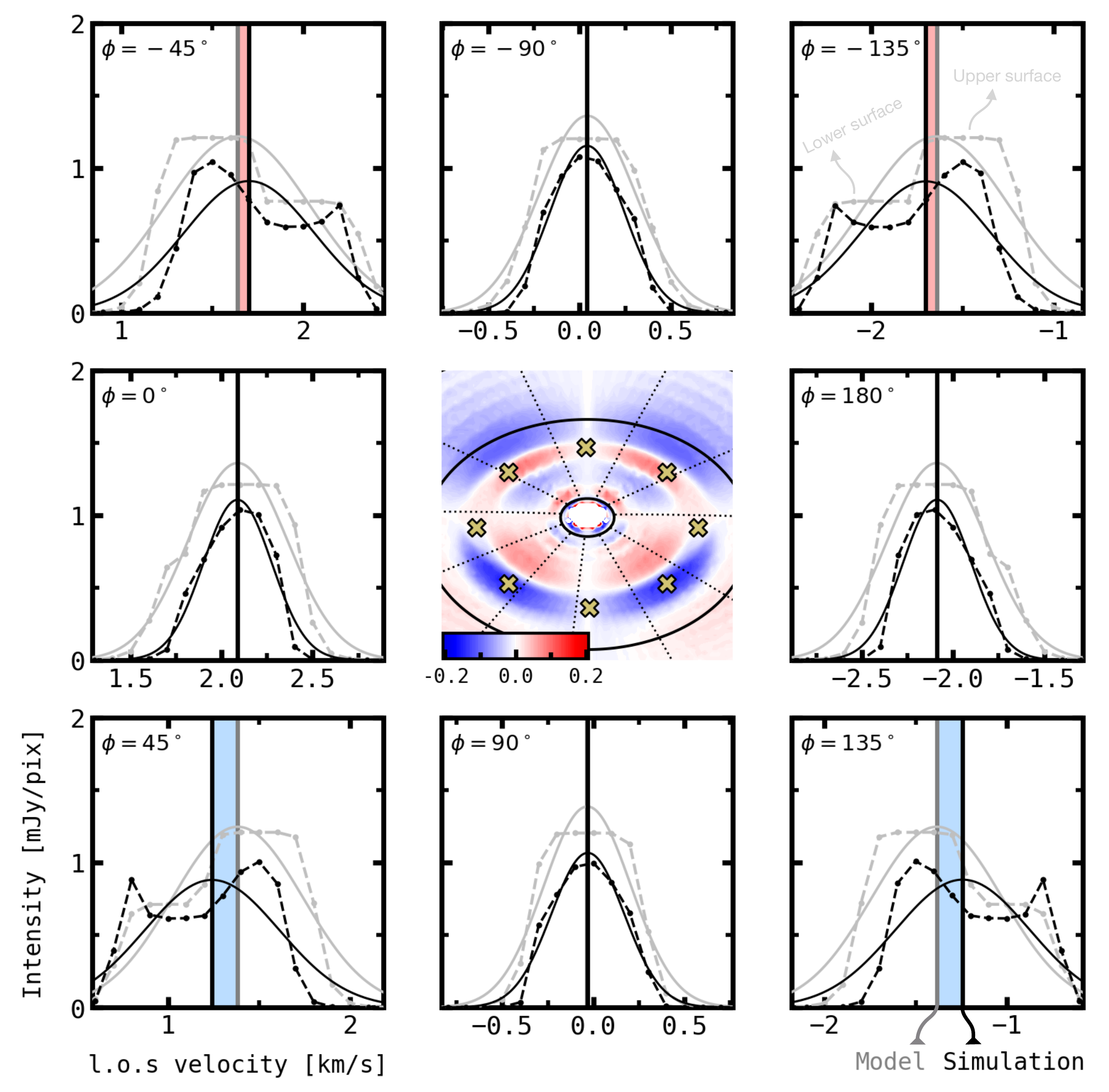}
     \caption{Comparison between simulation (dashed black) and model intensity profiles (dashed grey) extracted from different azimuths in the gap. Asymmetric line profiles are due to the contribution of both upper and lower emitting surfaces as indicated in the top right panel, whereas symmetric profiles are shaped by the upper surface only. The solid black and grey lines are the Gaussian fits of the profiles and the vertical lines are the associated line centroids. The middle panel shows centroid velocity residuals, in km\,s$^{-1}$, zoomed-in on the central region of the Keplerian disc of Fig. \ref{fig:azimuth_residuals_noplanet}. The crosses are all at $R=100$\,au, and indicate the exact location of the intensity profiles in the surrounding panels. The gap was initially carved by a 1.0\,$\Mj$ planet; then the planet was removed and the gas velocity set to be fully Keplerian for this analysis. The model, on the other hand, does not contain a gap, and so it systematically overestimates peak intensities there. Even though the disc rotation is Keplerian, there are high centroid velocity residuals due to differences between the simulation and model intensities. This effect appears all over the gap where the line profiles are asymmetric due to the contribution of the lower surface of the disc. 
              }
              
         \label{fig:gaussian_profiles_gap}
\end{figure*}

Our analysis indicates that gaps and emitting surfaces must be taken into account to fully understand kinematical observables. In particular, the gap and the uneven intensity contribution of the upper and lower surfaces of the disc can trigger artificial deviations from Keplerian rotation, which might appear similar to the perturbations on the gap edges (see Fig. \ref{fig:dens+temp}, right panel) but should not be confused with these latter. For instance, from Figure \ref{fig:azimuth_residuals_noplanet} we notice that even after forcing the gas motions to be Keplerian, our line centroid method overestimates(underestimates) velocities on the gap at negative(positive) azimuths of the disc. As illustrated in Figure \ref{fig:gaussian_profiles_gap}, this is due to the drop of intensity at the gap which abruptly alters the upper-to-lower surface intensity ratio, shifting the observed line centroids to non-Keplerian velocities. Another more direct observational consequence is that gaps alone are also able to produce kinks (see e.g. Fig. \ref{fig:channel_maps}, middle rows) which may lead empirical methods to false positive inference of planets.

A proper model of the gap intensity would help reproduce the Keplerian pattern of the gap and in turn would facilitate the extraction of velocity fluctuations within it. Unfortunately, even a simple prescription of a gap implies at least three more model parameters (gap location, width, and depth), making our main goal of detecting planet-driven perturbations unnecessarily complex. Alternatively, one could simply study gas velocities at peak intensities (namely on the upper surface of the disc) only, instead of using full intensity profiles as our method does. However, this approach affects the accuracy of the retrieved velocities and omits any information coming from the lower surface of the disc, which we proved to be crucial for extracting velocity perturbations of planets at certain azimuths (see Sect. \ref{sec:discussion_magnitude_pert} and Fig. \ref{fig:channel_pos_neg}). Instead, we propose a simpler solution. The contribution of the gap can be readily excluded from the analysis by exploiting its symmetry around the projected minor axis, which we do in Sect. \ref{subsec:centroid_folding} by subtracting line centroids on the east from those on the west side of the disc.
 
\begin{figure*}
   \centering
   \includegraphics[width=0.36\textwidth]{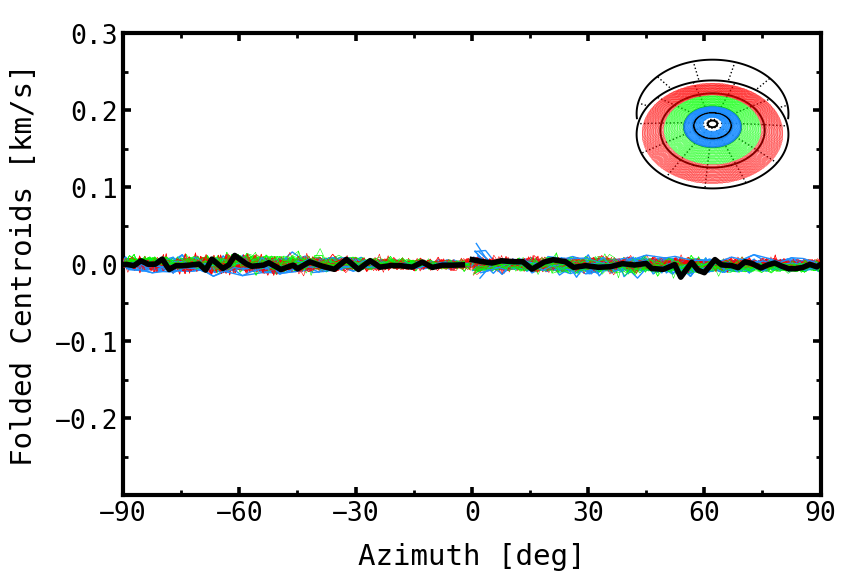}
   \includegraphics[width=0.6\textwidth]{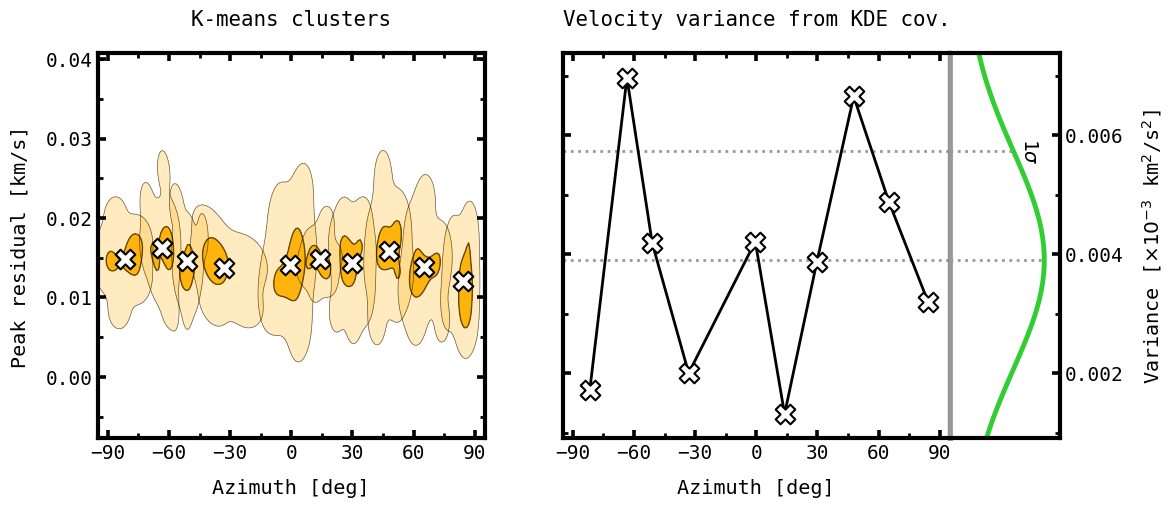}
     \caption{Same as Figs. \ref{fig:azimuthal_folded_centroids} and \ref{fig:clusters_1Mjup_PA45} but for azimuthally symmetric velocity perturbations. Unlike planet-driven perturbations, such a scenario leads to non-localised velocity fluctuations and hence to non-detections according to our Variance Peak method.
              }
         \label{fig:zonalflow_nondetection}
\end{figure*}

In addition, we conducted an experiment to study the impact of non-planet-induced zonal flows on our detection analysis. To do this, instead of forcing the disc velocity field to be fully Keplerian, we smoothed it out by taking azimuthally averaged velocities across all pixels. This step eliminates any localised planet perturbation while keeping high-velocity fluctuations at the gap edges. We then applied the same analysis detailed in Sect. \ref{sec:results}. Such a scenario effectively leads to non-detection of planets, with both azimuthal and radial peak velocity variances barely reaching a level of 1$\sigma$ (see Fig. \ref{fig:zonalflow_nondetection}). Again, this is because our line centroid folding procedure cancels any contribution from an azimuthally symmetric velocity field, regardless of its origin.   

%-----------------------------------------------------------------

%-----------------------------------------------------------------

\section{Supporting figures} \label{sec:appendix_figures}

%-----------------------------------------------------------------
\begin{figure*}
   \centering
   \includegraphics[width=1.0\textwidth]{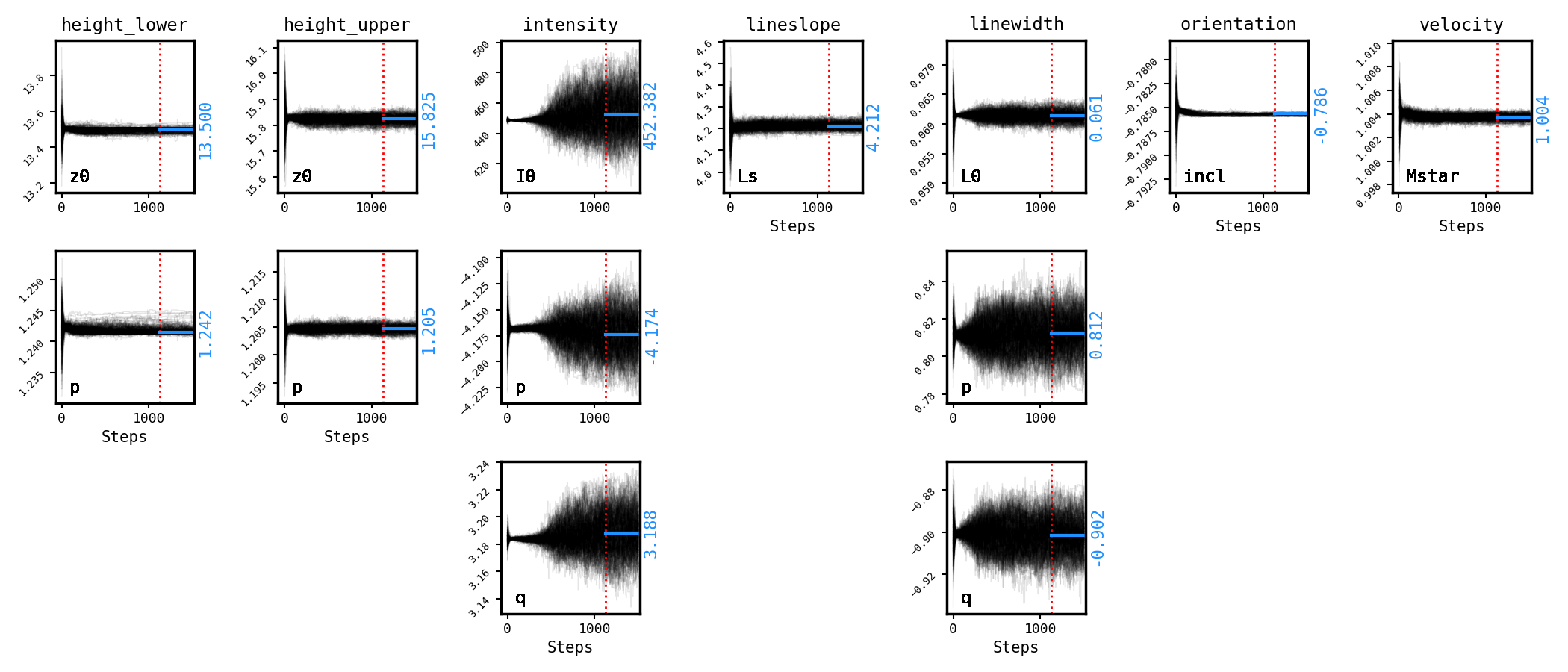}
     \caption{Converging parameter walkers obtained with the \discminer{} for the 0.3\,$\Mj$ snapshot, using 256 walkers and 1500 steps. This execution is preceded by an initial 1000-step run which is useful to find the seeding parameters before convergence. The dashed red line highlights the last quarter of walkers whose median corresponds to the reported best-fit parameters (in blue). 
              }
         \label{fig:walkers}
\end{figure*} 

%-----------------------------------------------------------------
\begin{figure*}
\centering
   \includegraphics[width=1.0\columnwidth]{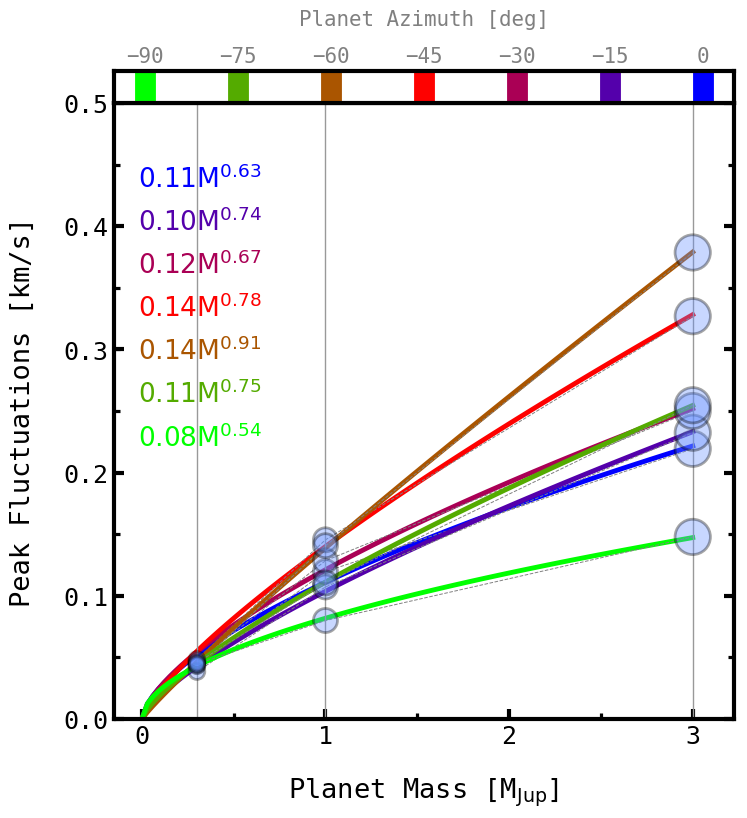} \hspace{0.4cm} 
   \includegraphics[width=0.928\columnwidth]{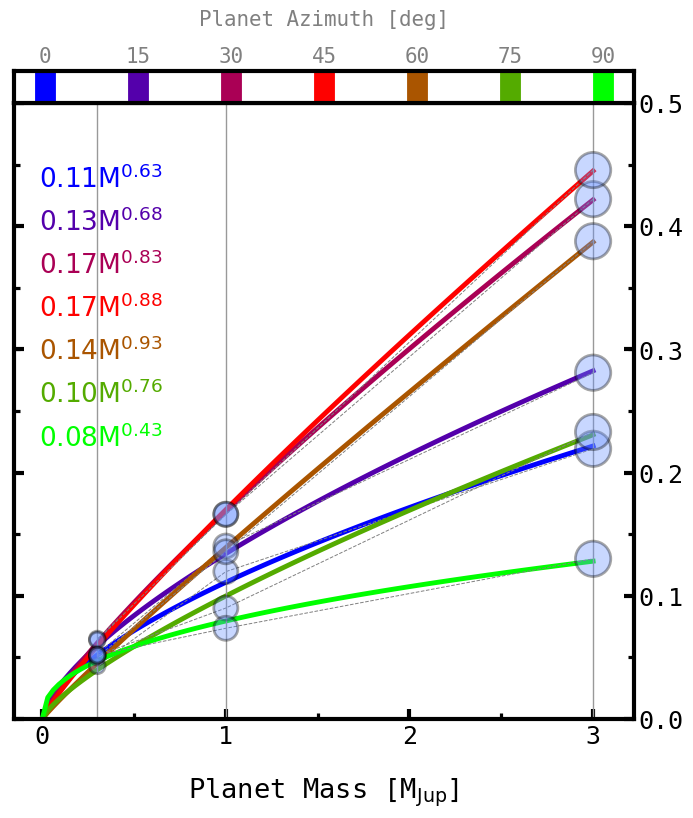}
   
     \caption{Peak velocity fluctuations against planet mass for different planet azimuths (coloured lines, \textit{left}: negative, \textit{right}: positive azimuths). A power-law fit of the form $\delta\upsilon=aM^b$ is shown in colour according to each planet azimuth.
              }
         \label{fig:fitmass}
\end{figure*} 
%-----------------------------------------------------------------

\end{appendix}
\end{document}